\documentclass[a4paper,12pt]{article}
\usepackage{jcappub} 
\usepackage{graphicx}
\usepackage{amsfonts}
\usepackage{amssymb,amsmath}
\usepackage{latexsym}
\usepackage{lscape}
\usepackage{color}
\usepackage{bm}
\usepackage{ulem}
\usepackage[a4paper]{geometry}
\input{colordvi.tex}

\newcommand{\siml}{\lesssim}
\newcommand{\simg}{\gtrsim}

\title{Probing the Universe through the Stochastic Gravitational Wave Background}

\author[a,b]{Sachiko Kuroyanagi}
\author[c]{Takeshi Chiba}
\author[d]{Tomo Takahashi}

\affiliation[a]{Department of Physics, Nagoya University, Chikusa, Nagoya 464-8602, Japan}
\affiliation[b]{Institute for Advanced Research, Nagoya University, Chikusa, Nagoya 464-8602, Japan}
\affiliation[c]{Department of Physics, College of Humanities and Sciences, Nihon University, Tokyo 156-8550, Japan}
\affiliation[d]{Department of Physics, Saga University, Saga 840-8502, Japan}

\emailAdd{skuro@nagoya-u.jp}
\emailAdd{chiba@phys.chs.nihon-u.ac.jp}
\emailAdd{tomot@cc.saga-u.ac.jp}

\abstract{Stochastic gravitational wave backgrounds, predicted in many
  models of the early universe and also generated by various
  astrophysical processes, are a powerful probe of the Universe.  The
  spectral shape is key information to distinguish the origin of the
  background since different production mechanisms predict different
  shapes of the spectrum.  In this paper, we investigate how precisely
  future gravitational wave detectors can determine the spectral shape
  using single and broken power-law templates.  We consider the
  detector network of Advanced-LIGO, Advanced-Virgo and KAGRA and the
  space-based gravitational-wave detector DECIGO, and estimate the
  parameter space which could be explored by these detectors.  We find
  that, when the spectrum changes its slope in the frequency range of
  the sensitivity, the broken power-law templates dramatically improve
  the $\chi^2$ fit compared with the single power-law templates
  and help to measure the shape with a good precision.  }

\begin{document}

\maketitle

\section{Introduction}

Gravitational waves (GWs) would have been generated in the course of
the evolution of the Universe from the very early era to the present.
Since GWs can penetrate through space without attenuation, they carry
invaluable information on phenomena in the very early Universe and
astrophysical processes, which cannot be unraveled by other
observations.

One such example is inflation, in which GWs as well as density 
perturbations are generated from quantum fluctuations
\cite{Starobinsky:1979ty,Rubakov:1982df}.  There are many other
possible sources of GWs from the early Universe, such as first-order
phase transition \cite{Kosowsky:1992rz,Kosowsky:1991ua,Kosowsky:1992vn,
Kamionkowski:1993fg,Hindmarsh:2013xza}, preheating after inflation
\cite{Khlebnikov:1997di}, topological defects
\cite{Vilenkin:1981zs,Vachaspati:1984gt,Damour:2000wa,Krauss:1991qu,Fenu:2009qf},
and so on. These GWs are considered as those from uncorrelated and
unresolved sources and generate a stochastic background of
GWs. Furthermore, various stochastic GW backgrounds of astrophysical
origin have been discussed, such as binaries of compact objects (black
holes, neutron stars, white dwarfs)
\cite{Meacher:2015iua,Farmer:2003pa}, stellar core collapse
\cite{Buonanno:2004tp,Crocker:2017agi}, r-mode instability of neutron
stars \cite{Ferrari:1998jf}, magnetars \cite{Regimbau:2005ey} and so
on. The detection of such stochastic GW backgrounds would give us an
important insight on cosmology and astrophysics.  In fact, the
world-wide detector network of Advanced-LIGO (aLIGO), Advanced-Virgo
(aVirgo) and KAGRA will increase the sensitivity of the GW background
up to $\Omega_{\rm GW}\sim 10^{-9}$ at the frequency of $10-100$~Hz.
In addition, the future space-based gravitational-wave detector
Deci-Hertz Interferometer Gravitational-wave Observatory (DECIGO)
\cite{Seto:2001qf,Kawamura:2011zz} might be able to detect stochastic
GWs up to $\Omega_{\rm GW}\sim 10^{-16}$ at the frequency of $0.1 -
1$~Hz.

Since there are a lot of possible sources of stochastic GW backgrounds
of various origins, we should prepare for its future detection. As
described in Sec. 2, most of the spectra of stochastic GW backgrounds
cannot be fitted by a single power-law as usually assumed but, rather,
by a broken power-law, which can be characterized by two spectral
indices, peak frequency, and amplitude.  The spectral shape contains
information on the source of the background, hence accurate modeling
of the spectral shape would help to uncover the origin and the nature
of this source.  Fitting the stochastic background well-described by a
broken power-law spectrum using a single power-law template would lead
to a biased estimate of the spectral index, and useful information of
the source would be lost.  In this paper, focusing on the future
detector network of aLIGO-aVirgo-KAGRA and the next-generation GW
detectors such as DECIGO, we investigate how accurately we can extract
the information on the parameters of the broken power-law
templates from measurements of the spectrum of stochastic GW
background (see \cite{Bose:2005fm}, for an estimation of the number of
templates required in the LIGO experiment in the stochastic GW
background search with a broken power-law fit).

The organization of this paper is as follows.  In
Sec.~\ref{sec:GW_source}, we review the sources of the stochastic GW
background (cosmological ones in Sec.~\ref{subsec:cosmological} and
astrophysical ones in Sec.~\ref{subsec:astrophysical}) and list the
quantities characterizing the GW spectrum such as the amplitude, the
spectral index and the frequency.  In Sec.~\ref{sec:methodology}, we
describe the method of the analysis to obtain expected constraints
from future observations mainly by adopting the Fisher matrix and
demonstrate how the parameter estimation is biased when we use an
unsuitable template. In Sec.~\ref{sec:result}, we forecast the
expected constraints on the parameters by the future detector network
of aLIGO-aVirgo-KAGRA and the next-generation GW detector
DECIGO. Sec.~\ref{sec:summary} is devoted to summary.

\section{GW  sources}
\label{sec:GW_source}

In this section, we summarize (possible) GW sources of cosmological
and astrophysical origins, which have been suggested in the
literature.  Here we do not intend to set a thorough list, but we discuss
the sources which have been investigated relatively well.

GWs are described by the tensor perturbation $h_{ij}$ in the
Friedmann-Robertson-Walker (FRW) spacetime:
\begin{equation}
ds^2 = - dt^2 + a^2 (t) \left( \delta_{ij} + h_{ij} \right) dx^i dx^j,
\end{equation}
with $a(t)$ being the scale factor of the Universe. Here we consider a
flat Universe and $h_{ij}$ satisfying the transverse-traceless
condition: $\partial^i h_{ij} = h^i_{~i}=0$.  The energy density of
the GWs is given by
\begin{equation}
\rho_{\rm GW}  = \frac{1}{64 \pi G} \left\langle (\partial_t  h_{ij})^2 + \left( \frac{\nabla}{a} h_{ij} \right)^2 \right\rangle,
\end{equation}
where the bracket describes the spatial average.

To characterize the spectral amplitude of GWs, we use the dimensionless
quantity $\Omega_{\rm GW}$, which describes the energy density of GWs per
logarithmic interval of the frequency $f$ at the present time,
normalized by the critical density $\rho_{\rm crit} = 3 H_0^2
/(8\pi G)$:
\begin{equation}
\Omega_{\rm GW} \equiv  \frac{1}{\rho_{\rm crit}} \frac{ d \rho_{\rm GW}}{ d \ln f}.
\end{equation}
One may approximate the GW spectrum using a broken power-law as
\begin{equation}
\label{eq:GW_spectrum}
\Omega_{\rm GW} (f)  =
\begin{cases}
      &  \Omega_{\rm GW \ast} \left( \displaystyle\frac{f}{f_\ast} \right)^{n_{\rm GW1}}   \qquad \text{for}~~ f < f_\ast, \\ \\
      & \Omega_{\rm GW \ast} \left( \displaystyle\frac{f}{f_\ast} \right)^{n_{\rm GW2}}   ~~\qquad \text{for}~~ f > f_\ast, 
\end{cases}
\end{equation}
where $\Omega_{\rm GW *}$ is the amplitude at $f=f_\ast$ (the peak
frequency or the reference frequency) with $f_\ast$ being the frequency at which the spectral
dependence changes, and $n_{\rm GW1}$ and $n_{\rm GW2}$ are the
spectral index for $f < f_\ast$ and $f > f_\ast$, respectively.
Although not all the models can  well be described by this simple
form, in the following, we provide typical values of $\Omega_\ast,
f_\ast, n_{\rm GW1}$ and $n_{\rm GW2}$ for various cosmological and
astrophysical stochastic backgrounds.

\subsection{Cosmological sources}
\label{subsec:cosmological}

First, we list cosmological sources. See also
\cite{Binetruy:2012ze,Caprini:2018mtu} for a collection of some
cosmological sources.  All the models we describe in this subsection
are summarized in Table \ref{table:cosmological}.

\bigskip
\noindent
$\bullet$ {\bf First-order phase transition} \vspace{2mm} \\
It has been argued that significant GWs can be generated during
first-order phase transition in the early universe (for example, an
electroweak-scale phase transition
\cite{Apreda:2001us,Grojean:2006bp}).  The GW spectrum depends on the
mechanisms taking place during the phase transition.  There are three
processes generating GWs: bubble collision, turbulence and sound
waves.  Below, we quote the spectral indices, the peak frequency and
the amplitude of the GW spectrum from these processes separately
\footnote{
There have been some works discussing the discrimination of models of phase transition by using GW spectrum \cite{Jinno:2017ixd,Croon:2018erz}. 
}.

\bigskip
\noindent
(i) Bubble collision \cite{Kosowsky:1992rz,Kosowsky:1992vn,Caprini:2007xq,Huber:2008hg,Caprini:2015zlo} \vspace{2mm} \\
In a first-order phase transition, bubbles are nucleated.  They
rapidly expand and collide, sourcing a large amount of GWs.  The
GWs from bubble collision has spectral indices 
\begin{equation}
n_{\rm GW1} = 2.8,
\qquad 
n_{\rm GW2} = -1,
\end{equation}
The peak frequency of the GWs generated at the time of phase transition is written as 
\begin{equation}
f_{\rm PT} = \beta \left( \displaystyle\frac{0.62}{1.8 - 0.1 v_w + v_w^2} \right),
\end{equation}
with $v_w$ being the bubble wall velocity.  When it is redshifted to
the present-day frequency, we have
\begin{equation}
f_\ast \sim 10^{-5}  \left( \displaystyle\frac{f_{\rm PT}}{\beta} \right) \left( \displaystyle\frac{\beta}{H_{\rm PT}}  \right)  \left( \displaystyle\frac{T_{\rm PT}}{100~{\rm GeV}} \right)
\quad{\rm [Hz]},
\end{equation}
where $ \beta \simeq \dot{\Gamma}/\Gamma$ with $\Gamma$ being the
bubble nucleation rate, and $H_{\rm PT}$ and $T_{\rm PT}$ are the
Hubble rate and the temperature at the time of the phase transition.
The amplitude at peak frequency today is given by
\begin{equation}
\Omega_{\rm GW \ast} \sim 10^{-5}  \left( \displaystyle\frac{H_{\rm PT}}{\beta}  \right)^2  \left( \displaystyle\frac{\kappa_\phi \alpha}{1+\alpha}  \right)^2   \left( \displaystyle\frac{0.11 v_w^3}{0.42 + v_w^2}  \right),
\end{equation}
where $ \kappa_\phi$ is the fraction of the vacuum energy converted
into the gradient energy of a scalar field.  For analytic
calculations, see
\cite{Caprini:2007xq,Caprini:2009fx,Jinno:2016vai,Jinno:2017fby}.

\bigskip
\noindent
(ii) Turbulence  \cite{Kamionkowski:1993fg,Binetruy:2012ze,Caprini:2009yp,Caprini:2015zlo} \vspace{2mm} \\
Subsequent magnetohydrodynamic (MHD) turbulent cascades after bubble collisions also source
GWs.  The spectral indices, the peak frequency and the present-day
amplitude can be written as
\begin{equation}
n_{\rm GW1} = 3,
\qquad 
n_{\rm GW2} = -\frac53,
\end{equation}

\begin{equation}
f_\ast \sim 3 \times 10^{-5}  \left( \displaystyle\frac{1}{v_w}  \right)  \left( \displaystyle\frac{\beta}{H_{\rm PT}}  \right)  \left( \displaystyle\frac{T_{\rm PT}}{100~{\rm GeV}} \right) 
\quad{\rm [Hz]},
\end{equation}

\begin{equation}
\Omega_{\rm GW \ast} \sim 3 \times 10^{-4}  \left( \displaystyle\frac{H_{\rm PT}}{\beta}  \right)  \left( \displaystyle\frac{\kappa_{\rm turb} \alpha}{1+\alpha}  \right)^{3/2}    v_w,
\end{equation}
where $\kappa_{\rm turb}$ is the fraction of latent heat converted into turbulence.

\bigskip
\noindent
(iii) Sound waves \cite{Hindmarsh:2013xza,Caprini:2015zlo,Hindmarsh:2015qta} \vspace{2mm} \\
Sound waves in the plasma fluid are also an important source of GWs.
For the case of sound waves, $n_{\rm GW1}, n_{\rm GW2}, f_\ast$ and
$\Omega_{\rm GW \ast}$ are given by
\begin{equation}
n_{\rm GW1} = 3,
\qquad 
n_{\rm GW2} = - 4,
\end{equation}

\begin{equation}
f_\ast \sim 2 \times 10^{-5}  \left( \displaystyle\frac{1}{v_w}  \right)  \left( \displaystyle\frac{\beta}{H_{\rm PT}}  \right)  \left( \displaystyle\frac{T}{100~{\rm GeV}} \right) \quad{\rm [Hz]},
\end{equation}

\begin{equation}
\Omega_{\rm GW \ast} \sim 3 \times 10^{-6}  \left( \displaystyle\frac{H_{\rm PT}}{\beta}  \right)  \left( \displaystyle\frac{\kappa_v \alpha}{1+\alpha}  \right)^{2}    v_w,
\end{equation}
where $ \kappa_v$ is the fraction of latent heat converted into the
bulk motion of the fluid. For a recent study of GWs from sound waves,
see \cite{Hindmarsh:2017gnf}.

\bigskip
\bigskip
\noindent
$\bullet$ {\bf Preheating} \vspace{2mm} \\
During preheating stage, GWs can be generated from violent production
of particles via a parametric resonance (see
\cite{Allahverdi:2010xz,Amin:2014eta} for a recent review on
preheating), and there have been a lot of studies on the generation of
GWs from preheating (see \cite{Khlebnikov:1997di} for a pioneering
work).  Although a numerical simulation is needed to precisely
calculate the GW spectrum, here we describe some approximate (fitting)
formula for the GW spectrum.  Below, we present only the cases of
preheating into scalars, but preheating into gauge fields is also
studied in the literature
\cite{Dufaux:2010cf,Tranberg:2017lrx,Adshead:2018doq}.

\bigskip
\noindent
(i) Case with $V = \displaystyle\frac14 \lambda \phi^4$  \vspace{2mm} \\
First, let us consider a model with 
\begin{equation}
V (\phi, \chi) = \displaystyle\frac14 \lambda \phi^4 + \frac12 g \phi^2 \chi^2,
\end{equation}
where $\phi$ is the inflaton which decays into another scalar field
$\chi$.  Although a quartic chaotic inflation model is now ruled out
by Planck data, for reference, we fix the value of $\lambda$ to give
the right amplitude of primordial density fluctuations, i.e., $\lambda
\simeq 10^{-14}$.  In this model, the spectral indices are given by
\begin{equation}
n_{\rm GW1} = 3,
\qquad 
n_{\rm GW2} = {\rm cutoff}.
\end{equation}
At higher frequency, the GW spectrum decays exponentially and cannot
be well fitted by a constant power law.  From now on, we denote such a case
as ``cutoff''.  Once we fix the value of $\lambda$, i.e, the
inflation scale, the peak frequency is approximately fixed as
\cite{Figueroa:2017vfa}
\begin{equation}
f_\ast \sim 10^7  \quad{\rm [Hz]}.
\end{equation}
Ref.~\cite{Figueroa:2017vfa} has shown that the peak amplitude can be
fitted to the so-called resonance parameter $q(= g^2/\lambda)$.  Since
the amplitude oscillates depending on $q$, we give a range in the
formula below:
\begin{equation}
3.4 \times 10^{-12} \left( \frac{q}{100} \right)^{-0.42}
<
\Omega_{\rm GW \ast}  h^2 
<
2.4 \times 10^{-11} \left( \frac{q}{100} \right)^{-0.56},
\end{equation}
where $h$ is the reduced Hubble constant.
Thus, roughly speaking, the peak amplitude is 
\begin{equation}
\Omega_{\rm GW \ast}  \sim 10^{-11}  \left( \frac{q}{100} \right)^{-0.5}.
\end{equation}

\bigskip
\noindent
(ii) Hybrid
\cite{GarciaBellido:2007dg, Dufaux:2008dn} \vspace{2mm} \\
For a hybrid-type inflationary model,  the potential can be given by
\begin{equation}
V = \frac14 \lambda \left( \sigma^2  - v^2 \right)^2 + \frac12 g^2 \phi^2 \sigma^2 + V_{\rm inf} (\phi),
\end{equation}
where $\phi$ is the inflaton, $V_{\rm inf} (\phi)$ is its potential
controlling the inflationary dynamics during inflation, $\lambda$ and
$g$ are coupling constants, and $v$ is the VEV of a field $\sigma$.
But we do not need to specify it here since the generation of GWs from
preheating does not depend on the details of the potential during
inflation.  For this model, the spectral indices, the peak frequency
and the amplitudes can be roughly given as, for the case of $g^2/
  \lambda \ll 1$,
\begin{equation}
n_{\rm GW1} = 2,
\qquad 
n_{\rm GW2} = {\rm cutoff},
\end{equation}
\begin{equation}
f_\ast \sim \frac{g}{\sqrt{\lambda}} \lambda^{1/4} 10^{10.25} \quad{\rm [Hz]},
\end{equation}

\begin{equation}
\Omega_{\rm GW \ast} \sim 10^{-5} \left( \frac{\lambda}{g^2} \right)^{1.16} \left( \frac{v}{M_{\rm pl}} \right)^2.
\end{equation}

\bigskip
\bigskip
\noindent
$\bullet$ {\bf Cosmic strings } \cite{Vilenkin:1981zs,Vachaspati:1984gt} \vspace{2mm} \\
Cosmic strings are one-dimensional topological defects, which arise
naturally in field theories, as well as in inflationary scenarios
based on superstring theory. They are known to emit strong GW bursts
from pathological structures, such as cusps and kinks
\cite{Damour:2000wa}, during their evolution.  When GWs from all the
strings are numerous, their signals overlap and become a stochastic GW
background.

\bigskip
\noindent
(i) Loops 1
\cite{Damour:2001bk,Damour:2004kw,Siemens:2006yp,DePies:2007bm,Olmez:2010bi,Sanidas:2012ee,Sanidas:2012tf,Kuroyanagi:2012wm,Kuroyanagi:2012jf,Binetruy:2012ze,Sousa:2013aaa,Blanco-Pillado:2017oxo,Ringeval:2017eww} \vspace{2mm} \\
Cosmic string loops are known to generate a GW background at high
frequencies.  The loops formed in the late matter-dominated era give
rise to a GW background with a peak-like shape.  Taking into account
the uncertainties in the string network modeling, the spectral indices
roughly range as
\begin{equation}
n_{\rm GW1} = [1,2],
\qquad 
n_{\rm GW2} = [-1,-0.1],
\end{equation}
with
\begin{equation}
f_\ast  \sim 3\times 10^{-8}\left( \frac{G\mu}{10^{-11}} \right)^{-1}
\quad{\rm [Hz]},
\end{equation}
where $G$ is the gravitational constant and $\mu$ is the string
tension.  Note that this dependence holds only for $\alpha_{\rm loop}
\gg \Gamma G\mu$ where $\Gamma$ characterizes GW emission efficiency
and $\alpha_{\rm loop}$ is the typical initial size of loops
normalized with respect to the loop formation time $t_i$.
When $\alpha_{\rm loop} \gg \Gamma G\mu$, the dependence is
\begin{equation}
f_\ast  \sim 3\times 10^{-8}\left(\frac{\alpha_{\rm loop}}{10^{-9}} \right)^{-1}
\quad{\rm [Hz]},
\end{equation}
The amplitude strongly depends on the string parameters such as
tension $G\mu$ and initial loop size $\alpha_{\rm loop}$.  When one
considers $\alpha_{\rm loop} \gg \Gamma G\mu$, the amplitude at peak is
roughly given by
\begin{equation}
  \Omega_{\rm GW \ast} \sim 10^{-9}
  \left(\frac{G\mu}{10^{-12}}\right)
  \left(\frac{\alpha_{\rm loop}}{10^{-1}}\right)^{-1/2}.
\end{equation}
For $\alpha_{\rm loop} \ll \Gamma G\mu$, the parameter dependence becomes
\begin{equation}
  \Omega_{\rm GW \ast} \sim 10^{-12}
  \left(\frac{G\mu}{10^{-12}}\right).
\end{equation}
Note that, in the case of cosmic superstrings, the reconnection
probability also affects the amplitude.

\bigskip
\noindent
(ii) Loops 2
\cite{Damour:2001bk,Damour:2004kw,Siemens:2006yp,DePies:2007bm,Olmez:2010bi,Sanidas:2012ee,Sanidas:2012tf,Kuroyanagi:2012wm,Kuroyanagi:2012jf,Binetruy:2012ze,Sousa:2013aaa,Blanco-Pillado:2017oxo,Ringeval:2017eww} \vspace{2mm} \\
At higher frequencies, GWs from loops formed during the
radiation-dominated era are the dominant contribution and the spectrum
becomes flat.  Thus, around the intermediate frequency where we see
both contributions from loops formed in the radiation-dominated and
the matter-dominated phases, the spectral indices change as
\begin{equation}
n_{\rm GW1} = [-1,-0.1],
\qquad 
n_{\rm GW2} = 0.
\end{equation}
The transition frequency and the amplitude strongly depends on the modeling of the cosmic string network and the calculation method, but the rough expectation is 
\begin{equation}
f_\ast \sim 3\times 10^{-5}\left( \displaystyle\frac{G\mu}{10^{-11}} \right)^{-1}
\quad{\rm [Hz]},
\end{equation}
\begin{equation}
  \Omega_{\rm GW \ast} \sim 10^{-9.5} \left(\frac{G\mu}{10^{-12}} \right)
  \left( \frac{\alpha_{\rm loop}}{10^{-1}} \right)^{-1/2},
\end{equation}
for $\alpha_{\rm loop} \gg \Gamma G\mu$, and 
\begin{equation}
f_\ast \sim 3\times 10^{-3}\left(\displaystyle\frac{\alpha_{\rm loop}}{10^{-9}}\right)
^{-1}\quad{\rm [Hz]},
\end{equation}
\begin{equation}
  \Omega_{\rm GW \ast} \sim 10^{-14}
  \left(\frac{G\mu}{10^{-12}}\right),
\end{equation}
for $\alpha_{\rm loop} \ll \Gamma G\mu$.

\bigskip
\noindent
(iii) Infinite strings \cite{Kawasaki:2010yi,Matsui:2016xnp} \vspace{2mm} \\
Kinks on infinite strings generate a GW background over all
frequencies. Typically, the amplitude is smaller than the one from
loops, but it becomes important at low frequencies where loops do not
emit GWs.  The spectral index slightly depends on the expansion
rate of the Universe when kinks are generated, but typically the
spectrum is almost flat.  When combined with the GWs from loops which
produce GWs at high frequencies, one may find a break in the spectrum
such as
\begin{equation}
n_{\rm GW1} =[0,0.2],
\qquad 
n_{\rm GW2} = [1,2],
\end{equation}
The transition frequency is highly model dependent since parameter
dependencies of GW spectra from loops and infinite strings are
different, and hence we do not set a value for $f_\ast$.
For a typical parameter choice, the GW amplitude can be roughly given by 
\begin{equation}
\Omega_{\rm GW \ast}  \sim  10^{-12} \left(\frac{G\mu}{10^{-8}} \right).
\end{equation}
The prefactor can vary depending on the transition frequency, but it
should be in the range of $ [ 10^{-11}, 10^{-13}] $ for $G \mu =
10^{-8}$.

\bigskip
\bigskip
\noindent
$\bullet$ {\bf Domain walls }
\cite{Vilenkin:1981zs,Gleiser:1998na,Hiramatsu:2013qaa,Kawasaki:2011vv,Saikawa:2017hiv} \vspace{2mm} \\
The existence of domain walls is in conflict with cosmological
observations, since their energy density easily dominates that of the
universe.  However, this problem can be avoided by considering
unstable domain walls and their annihilation in the early Universe may
produce a significant amount of gravitational waves.

Numerical simulations
\cite{Hiramatsu:2013qaa,Kawasaki:2011vv,Saikawa:2017hiv} find the
spectral dependencies of the power spectrum as
\begin{equation}
n_{\rm GW1} =3,
\qquad 
n_{\rm GW2} = -1,
\end{equation}
with  typical frequency 
\begin{equation}
f_\ast  \sim 10^{-9}\left(\displaystyle\frac{T_{\rm ann}}{10^{-2}{\rm GeV}}\right)
\quad{\rm [Hz]},
\end{equation}
where $T_{\rm ann}$ is the temperature of the universe at domain wall
annihilation.
The amplitude is determined by the domain wall tension $\sigma$ as
\begin{equation}
\Omega_{\rm GW \ast}  \sim 10^{-17}\left(\displaystyle\frac{\sigma}{1{\rm TeV}^3}\right)^2\left(\displaystyle\frac{T_{\rm ann}}{10^{-2}{\rm GeV}}\right)^{-4}.
\end{equation}

\bigskip
\bigskip
\noindent
$\bullet$ {\bf  Self-ordering scalar fields} \cite{Krauss:1991qu} \vspace{2mm} \\
A phase transition which breaks global O($N$) symmetry of scalar fields
generates a spatial gradient of the scalar fields on superhorizon
scales, because each causally disconnected region of the Universe gets
arbitrarily different directions of the fields.  When the modes re-enter
the horizon, the fields release gradient energy by the self-ordering
of the Nambu-Goldstone modes, and they continuously
source GWs at the horizon scale. \\

\bigskip
\noindent
(i) Radiation-dominated phase
\cite{JonesSmith:2007ne,Fenu:2009qf,Giblin:2011yh,Figueroa:2012kw} \vspace{2mm} \\
If all the GW modes of interest enter the horizon during the
radiation-dominated phase, GWs have a scale-invariant spectrum at the
frequencies of interferometer experiments, and hence we have
\begin{equation}
n_{\rm GW1} =0,
\qquad 
n_{\rm GW2} = 0.
\end{equation}
Therefore, there is no well-defined peak frequency $f_\ast$ in this
model.  The spectral amplitude depends on the number of the scalar
field components $N$ and the VEV of the fields $v$,
\begin{equation}
\label{OGWself}
\Omega_{\rm GW \ast}   \sim \displaystyle\frac{511}{N} \Omega_{\rm rad} \left( \displaystyle\frac{v}{M_{\rm pl}} \right)^4.
\end{equation}

\bigskip
\noindent
(ii) Effect of reheating \cite{Kuroyanagi:2015esa} \vspace{2mm} \\
The frequency dependence of the GW spectrum is affected by the
expansion rate of the Universe.  If the expansion rate of the Universe
evolves like a matter-dominated phase during reheating, the spectral
indices are 
\begin{equation}
n_{\rm GW1} =0,
\qquad 
n_{\rm GW2} = -2,
\end{equation}
and the peak frequency can be written as 
\begin{equation}
f_\ast \sim 0.4 \left( \displaystyle\frac{T_R}{10^7~{\rm GeV}}\right)
\quad{\rm [Hz]},
\end{equation}
where $T_R$ is the temperature of the Universe when reheating is
completed.  The flat part $n_{\rm GW1}=0$ corresponds to the modes
which enter the horizon during the radiation-dominated phase after
reheating, and its amplitude is given as in Eq.~(\ref{OGWself}), while
the amplitude of the modes which enter during reheating is suppressed.
So the amplitude is
\begin{equation}
\Omega_{\rm GW \ast}  \sim \displaystyle\frac{511}{N} \Omega_{\rm rad} \left( \displaystyle\frac{v}{M_{\rm pl}} \right)^4.
\end{equation}

\bigskip
\bigskip
\noindent
$\bullet$ {\bf  Magnetic field}  \cite{Caprini:2001nb,Caprini:2006jb} \vspace{2mm} \\
Magnetic fields are considered to be present at almost all scales in
the Universe. In particular, they exist even in the intergalactic
medium \cite{Neronov:1900zz}, which may have originated in the early
Universe.  It has been argued that such primordial magnetic fields can
arise from inflation \cite{Turner:1987bw,Ratra:1991bn}, phase
transition \cite{Vachaspati:1991nm,Enqvist:1993np} and so on.  Here we
describe the GW power spectrum having magnetic fields generated from
phase transition in mind.  The slope of the power spectrum is
predicted to be
\begin{equation}
n_{\rm GW1} =3 ,
\qquad 
n_{\rm GW2} = \alpha_B + 1,
\end{equation}
where we have assumed that the initial magnetic field power spectrum
is given by $P_B \propto k^2$ and $\propto k^{\alpha_B}$ for scales
larger and smaller than the correlation scale respectively
\cite{Caprini:2006jb}, and the correlation length and the horizon
scale at the generation time are identical.  The characteristic
frequency and the amplitude at which the slope of the GW spectrum
changes can be roughly estimated as
\begin{equation}
f_\ast \sim 10^{-6} \left( \frac{T_\ast}{100~{\rm GeV}} \right)
\quad{\rm [Hz]},
\end{equation}
\begin{equation}
\Omega_{\rm GW \ast}  \sim 10^{-16} \left( \frac{B}{10^{-10}~{\rm G}} \right),
\end{equation}
where $T_\ast$ corresponds to the temperature at the production of
magnetic fields and $B$ is the magnetic field magnitude today.

\bigskip
\bigskip
\noindent
$\bullet$ {\bf  Inflation+reheating}   \cite{Turner:1993vb} \vspace{2mm} \\
Inflation generates almost scale-invariant GWs originating from the
quantum fluctuations in spacetime.  The primordial tensor power spectrum is given by
\begin{equation}
{\cal P}_{\rm inf} = \frac{8}{M_{\rm pl}^2} \left( \frac{H_\ast}{2\pi} \right)^2,
\end{equation}
where $H_\ast$ is the Hubble parameter at the horizon exit during inflation and it is
almost constant in the standard slow-roll inflationary models.
The present-day GW spectrum can be given by using the transfer
function $T(k)$, which describes the evolution of GWs after inflation,
\begin{equation}
\Omega_{\rm GW} (k) = \frac{1}{12} \left( \frac{k}{aH} \right)^2 T^2(k) {\cal P}_{\rm inf} (k),
\end{equation}
where $k=2\pi f$ is the wavenumber.  The explicit form of $T(k)$ is
given in
\cite{Nakayama:2008wy,Nakayama:2009ce,Kuroyanagi:2011fy,Kuroyanagi:2014nba}.

In the standard scenario, the inflaton oscillates at the bottom of its
potential during reheating. In such phase, the Universe behaves as
matter-dominated one if the inflaton potential has a quadratic form
at its bottom.  For modes entering the horizon during the
matter-dominated epoch, the transfer function scales as $T(k) \propto
k^{-2}$.  On the other hand, the modes entering the horizon during the
radiation-dominated epoch is $T(k) \propto k^{-1}$.  Because of the
transition from the matter-dominated epoch to the radiation-dominated
one at the end of reheating, the spectrum has 
\begin{equation}
n_{\rm GW1} \sim 0,
\qquad 
n_{\rm GW2} = -2.
\end{equation}
More precisely, $n_{\rm GW1}$ can be given by $n_{\rm GW1} =
-2\epsilon$ with $\epsilon = - \dot{H}/H^2$ being the first slow-roll
parameter, where, however, $|n_{\rm GW1} | \ll {\cal O}(1)$ in
general.

The characteristic frequency $f_\ast$ corresponds to the mode which
enters the horizon at the time of reheating.  Therefore, it can be
given as a function of the reheating temperature as
\begin{equation}
f_\ast \sim 0.3 \left( \displaystyle\frac{T_R}{10^7~{\rm GeV}}\right)
\quad{\rm [Hz]}.
\end{equation}
By using the tensor-to-scalar ratio, the amplitude $\Omega_{\rm GW
  \ast}$ is given by
\begin{equation}
\Omega_{\rm GW \ast}  \sim 2 \times 10^{-17} \left( \displaystyle\frac{r}{0.01} \right).
\end{equation}

\newpage
\noindent
$\bullet$ {\bf  Inflation+kination} \cite{Peebles:1998qn,Giovannini:1998bp,Giovannini:1999bh,Giovannini:1999qj,Tashiro:2003qp,Giovannini:2008tm} \vspace{2mm} \\
In some scenarios of the early Universe, the radiation-dominated epoch
is preceded by the so-called kination epoch, in which the energy
density of the Universe is dominated by the kinetic energy of a scalar
field.  Examples of this type of model include quintessential
inflation \cite{Peebles:1998qn}.  During the kination epoch, the
Hubble expansion rate decreases as $H \propto a^{-3}$ (the energy
density of the scalar field scales as $\rho_\phi \propto a^{-6}$,
which gives the transfer function of $T (k) \propto k^2$. Therefore,
the spectral indices for the GW spectrum are given by
\begin{equation}
n_{\rm GW1} \sim 0,
\qquad 
n_{\rm GW2} = 1.
\end{equation}
We note that here again $n_{\rm GW1}$ is given by $n_{\rm GW1} =
-2\epsilon$ which is close to 0.  The characteristic frequency
corresponds to the mode entering the horizon at the end of the
kination epoch.  By denoting the temperature at this epoch by $T_{\rm
  kin}$, $f_\ast$ is given by
\begin{equation}
f_\ast   \sim 0.3 \left( \displaystyle\frac{T_{\rm kin}}{10^7~{\rm GeV}}\right)
\quad{\rm [Hz]}.
\end{equation}
The amplitude  is given in the same way as for the
inflation+reheating case:
\begin{equation}
\Omega_{\rm GW \ast}  \sim 2 \times 10^{-17} \left( \displaystyle\frac{r}{0.01} \right).
\end{equation}

\bigskip
\bigskip
\noindent
$\bullet$ {\bf Particle production during inflation }   \cite{Cook:2011hg,Senatore:2011sp,Bartolo:2016ami} \vspace{2mm} \\
It has been argued that large GWs can be produced from particle
production during inflation
\cite{Cook:2011hg,Senatore:2011sp,Bartolo:2016ami}.  Let us consider a model 
where the inflaton $\phi$ couples to a $U(1)$ gauge field $F_{\mu\nu}$ as\footnote{
The GW production in models with an axion-SU(2) gauge field coupling
has been studied in
\cite{Adshead:2013qp,Adshead:2013nka,Dimastrogiovanni:2016fuu,Fujita:2017jwq,Thorne:2017jft}.
GWs generated from particle production in a bouncing model has been
discussed in \cite{Ben-Dayan:2016iks}.
}
\begin{equation}
{\cal L} \supset - \frac{\phi}{4 f} F_{\mu\nu} \tilde{F}^{\mu\nu},
\end{equation}
with $f$ being a coupling constant with the dimension of mass.  
In this model, gauge quanta can be significantly produced, which sources the GWs.
The primordial GW power spectrum is given by the sum of the contributions
from the usual inflationary vacuum and the particle production ones, which can be written as \cite{Barnaby:2010vf,Sorbo:2011rz}
\begin{equation}
P_{\rm GW} (k) = \frac{2 H_\ast^2}{\pi^2 M_{\rm pl}^2} + 8.7 \times 10^{-8} \frac{H_\ast^4}{M_{\rm pl}^4} \frac{e^{4\pi\xi}}{\xi^6},
\end{equation}
where $\xi$ is defined by 
\begin{equation}
\xi \equiv \frac{d \phi /dt}{2fH_\ast}.
\end{equation}
On large scale (low frequency), the contribution from the usual inflationary vacuum dominates, while 
on small scales (high frequency), the one from the particle production does.
Therefore  the spectral indices are written as  \cite{Bartolo:2016ami}
\begin{equation}
n_{\rm GW1} = -2 \epsilon\sim 0,
\qquad 
n_{\rm GW2} = -4 \epsilon + (4\pi \xi - 6) (\epsilon - \eta).
\end{equation}
The transition frequency corresponds to the one at which the GW
spectrum from particle production gets dominated over the one from the usual
inflationary tensor mode. The GW production from particle production is sensitive to
the parameters in the model such as $\xi$, and thus the transition frequency is  
highly dependent on the model parameters (see e.g., \cite{Barnaby:2011qe}).

The amplitude at the transition frequency is given as the same as the one from the usual inflationary vacuum (assuming an almost scale-invariant 
GW spectrum on lower frequency region), and hence it can 
be written as 
\begin{equation}
\Omega_{\rm GW \ast} \sim 2 \times 10^{-17} \left( \displaystyle\frac{r}{0.01} \right).
\end{equation}

\bigskip
\bigskip
\noindent
$\bullet$ {\bf  2nd-order perturbations}
\vspace{2mm} \\
At 2nd order in the cosmological perturbation theory, the scalar,
vector and tensor modes cannot be separated and they could affect one
another.  Typically the amplitude of the GWs generated from 2nd-order
scalar perturbations in standard slow-roll inflation is very small
when one considers the radiation-dominated epoch after inflation
\cite{Baumann:2007zm}.  However, considering a different Hubble
expansion of the Universe \cite{Assadullahi:2009nf} or
non-scale-invariant scalar perturbations, we can expect a large GW
amplitude.  For the latter, we introduce only the case related to
primordial black hole (PBH) formation \cite{Saito:2008jc}, but the
existence of other fields such as curvaton \cite{Kawasaki:2013xsa} and
instability of the standard model Higgs \cite{Espinosa:2018eve} can
also induce GWs with large amplitude.

\bigskip
\noindent
(i) Early matter phase \cite{Assadullahi:2009nf} \vspace{2mm}  \\
In \cite{Assadullahi:2009nf,Kohri:2018awv}, it has been argued that
2nd-order scalar perturbations induce the tensor mode which might be
detectable in the future GW observations if the Universe went through
an early matter-dominated phase.  When we consider the scale invariant
spectrum of scalar perturbations, the spectral indices of GWs are
given by
\begin{equation}
n_{\rm GW1} =1,
\qquad 
n_{\rm GW2} =  \textrm{drop-off}.
\end{equation}
The spectrum drops off sharply at the scale corresponding to the end
of inflation.  Therefore here we denote $n_{\rm GW2}$ as ``drop-off''.
Note that these indices change depending on the spectral shape of the
primordial scalar perturbations \cite{Alabidi:2013wtp}.  The typical
frequency is given by the reheating temperature $T_{\rm reh}$ and the
energy scale of inflation $M_{\rm inf}$,
\begin{equation}
f_\ast \sim 7\times 10^5  \left( \displaystyle\frac{T_{\rm reh}}{10^9~{\rm GeV}} \right)^{1/3} \left( \displaystyle\frac{M_{\rm inf}}{10^{16}~{\rm GeV}} \right)^{2/3}
\quad{\rm [Hz]},
\end{equation}
and the amplitude is given by 
\begin{equation}
\Omega_{\rm GW \ast}   \sim 10^{-12}  \left( \displaystyle\frac{T_{\rm reh}}{10^9~{\rm GeV}} \right)^{-4/3} \left( \displaystyle\frac{M_{\rm inf}}{10^{16}~{\rm GeV}} \right)^{4/3}.
\end{equation}

\bigskip
\noindent
(ii)  Primordial black holes  \cite{Saito:2008jc} \vspace{2mm} \\
PBHs can form when density fluctuations with large amplitude are
generated by some mechanism and such large scalar fluctuations induce GWs
as a second-order effect as discussed above.  If we assume primordial
scalar fluctuations with a peak-like shape, we can approximate the power
spectrum as a delta function as follows:
\begin{equation}
{\cal P}_\zeta (k) = {\cal A}^2 \delta (\ln (k/k_p) ),
\end{equation}
where ${\cal A}^2$ and $k_p$ are the amplitude and the wavenumber at
the peak.  With this kind of sharp scalar fluctuations, we also expect
GW generation and the spectrum is
\begin{equation}
n_{\rm GW1} =2,
\qquad 
n_{\rm GW2} = \textrm{drop-off},
\end{equation}
The spectrum drops off at the scale corresponding to the peak of
scalar fluctuations, which is related to the mass of the PBH
$M_{\rm PBH}$ as
\begin{equation}
f_\ast \sim 4 \times 10^{-2} \left( \frac{M_{\rm PBH}}{10^{20}~{\rm g}} \right)^{-1/2}
\quad{\rm [Hz]}.
\end{equation}
The amplitude at the peak is 
\begin{equation}
  \Omega_{\rm GW \ast}  \sim 7 \times 10^{-9} \left( \frac{{\cal A}^2}{10^{-3}}\right)^2.
\end{equation}
Note that the shape of the GW spectrum is different in the cases where
the power spectrum of fluctuation is amplified in a broad range of
scales and cannot be approximated by a delta function \cite{Alabidi:2012ex}.

\bigskip
\bigskip
\noindent
$\bullet$ {\bf  Pre-Big-Bang }
\cite{Buonanno:1996xc,Mandic:2005bd,Gasperini:1992pa,Brustein:1995ah} \vspace{2mm} \\
In a string theory-inspired cosmological scenario, the so-called pre-big bang
model, a blue-tilted GW spectrum can be generated.  In particular, the
lower frequency part of the spectrum is blue-tilted, while on higher
frequency, it can be flat or red/blue-tilted (see
\cite{Gasperini:2016gre} for a recent update and the detailed
spectrum). The spectral index for lower and higher frequency parts is
given by \cite{Buonanno:1996xc}
\begin{equation}
n_{\rm GW1} =3,
\qquad 
n_{\rm GW2} = 3 - 2\mu,
\end{equation}
where $\mu$ describes the growth of the dilaton during stringy phase
and $\mu \ge 0$.  The transition frequency corresponds to the one at
which the mode crosses the horizon at the beginning of the string
phase, which can be regarded as a model parameter in this scenario and
there is no typical phase. However, it has been argued that this
frequency can be around the one where LISA or aLIGO are sensitive.
The amplitude of the GW spectrum is estimated as
\begin{equation}
\Omega_{\rm GW \ast} \sim 1.4 \times 10^{-6} \left( \frac{H_s}{0.15 M_{\rm pl}} \right)^4,
\end{equation}
where $H_s$ is the Hubble parameter during the stringy phase.

\clearpage
\thispagestyle{empty}
\newgeometry{left=0cm,bottom=5mm}

\begin{landscape}
\begin{table}
\begin{center}
  \caption{Cosmological GW sources  }
  \label{ }
  \vspace{2mm}
\begin{tabular}{r|ccccl}
\hline \hline
source & $n_{\rm GW1}$ & $n_{\rm GW2}$ & $f_\ast$ [Hz] & $\Omega_{\rm GW}$  &  \\ \hline 
Phase transition (bubble collision)  
& $2.8$ & $-2$  
&  $ \sim 10^{-5}  \left( \displaystyle\frac{f_{\rm PT}}{\beta} \right) \left( \displaystyle\frac{\beta}{H_{\rm PT}}  \right)  \left( \displaystyle\frac{T_{\rm PT}}{100~{\rm GeV}} \right)$ 
& $ \sim 10^{-5}  \left( \displaystyle\frac{H_{\rm PT}}{\beta}  \right)^2  \left( \displaystyle\frac{\kappa_\phi \alpha}{1+\alpha}  \right)^2   \left( \displaystyle\frac{0.11 v_w^3}{0.42 + v_w^2}  \right)$   & \\
Phase transition (turbulence)  
& $3$ &  $-5/3 $ 
& $\sim 3 \times 10^{-5}  \left( \displaystyle\frac{1}{v_w}  \right)  \left( \displaystyle\frac{\beta}{H_{\rm PT}}  \right)  \left( \displaystyle\frac{T_{\rm PT}}{100~{\rm GeV}} \right)$  
& $ \sim 3 \times 10^{-4}  \left( \displaystyle\frac{H_{\rm PT}}{\beta}  \right)  \left( \displaystyle\frac{\kappa_{\rm turb} \alpha}{1+\alpha}  \right)^{3/2}    v_w $  & \\
Phase transition (sound waves)  
& $3$  & $-4$  
& $\sim 2 \times 10^{-5}  \left( \displaystyle\frac{1}{v_w}  \right)  \left( \displaystyle\frac{\beta}{H_{\rm PT}}  \right)  \left( \displaystyle\frac{T_{\rm PT}}{100~{\rm GeV}} \right)$  
&  $ \sim 3 \times 10^{-6}  \left( \displaystyle\frac{H_{\rm PT}}{\beta}  \right)  \left( \displaystyle\frac{\kappa_v \alpha}{1+\alpha}  \right)^{2}    v_w $  & \\
Preheating ($\lambda \phi^4$)  &$3$   &  cutoff    & $\sim 10^{7}$ & $\sim 10^{-11}  \left( \displaystyle\frac{g^2 / \lambda}{100} \right)^{-0.5}$  &  \\
Preheating (hybrid)  & $2$   &  cutoff  
& $\sim \displaystyle\frac{g}{\sqrt{\lambda}} \lambda^{1/4} 10^{10.25} $
& $\sim 10^{-5} \left( \displaystyle\frac{\lambda}{g^2} \right)^{1.16} \left( \displaystyle\frac{v}{M_{\rm pl}} \right)^2 $&  \\
Cosmic strings (loops 1)  &  $[1,2]$ & $[-1,-0.1]$ & 
$\sim 3\times 10^{-8}\left( \displaystyle\frac{G\mu}{10^{-11}} \right)^{-1}$
&
$\sim 10^{-9}\left(\displaystyle\frac{G\mu}{10^{-12}}\right)\left(\displaystyle\frac{\alpha_{\rm loop}}{10^{-1}}\right)^{-1/2}$
{\scriptsize (for $\alpha_{\rm loop} \gg \Gamma G\mu$)}
& \\
Cosmic strings (loops 2)  &  $[-1,-0.1]$ & $0$ & 
$\sim 3\times 10^{-8}\left( \displaystyle\frac{G\mu}{10^{-11}} \right)^{-1}$
& 
$\sim 10^{-9.5} \left(\displaystyle\frac{G\mu}{10^{-12}} \right) \left( \displaystyle\frac{\alpha_{\rm loop}}{10^{-1}} \right)^{-1/2}$
{\scriptsize (for $\alpha_{\rm loop} \gg \Gamma G\mu$)}
& \\
Cosmic strings (infinite strings)  & $[0,0.2]$ & $[0,0.2]$ & --- & $\sim 10^{-[11,13]} \left(\frac{G\mu}{10^{-8}} \right)$ & \\
Domain walls   & 3 & -1 
& $\sim 10^{-9}\left(\displaystyle\frac{T_{\rm ann}}{10^{-2}{\rm GeV}}\right)$ 
&  $\sim 10^{-17}\left(\displaystyle\frac{\sigma}{1{\rm TeV}^3}\right)^2\left(\displaystyle\frac{T_{\rm ann}}{10^{-2}{\rm GeV}}\right)^{-4}$  &\\
Self-ordering scalar fields & $0$  & $0$  & --- & $\sim \displaystyle\frac{511}{N} \Omega_{\rm rad} \left( \displaystyle\frac{v}{M_{\rm pl}} \right)^4$   & \\
Self-ordering scalar  + reheating & $0$  & $-2$  & $\sim 0.4 \left( \displaystyle\frac{T_R}{10^7~{\rm GeV}}\right)$  
& $\sim \displaystyle\frac{511}{N} \Omega_{\rm rad} \left( \displaystyle\frac{v}{M_{\rm pl}} \right)^4$ & \\
Magnetic fields &  $3$   & $\alpha_B +1$ & $\sim 10^{-6}\left(\displaystyle\frac{T_\ast}{10^{2}{\rm GeV}}\right)$ & $\sim 10^{-16}\left(\displaystyle\frac{B}{10^{-10}{\rm G}}\right)$ & \\ 
Inflation+reheating  & $\sim 0$  & $-2$ &  $\sim 0.3 \left( \displaystyle\frac{T_R}{10^7~{\rm GeV}}\right)$ &  $\sim 2 \times 10^{-17} \left( \displaystyle\frac{r}{0.01} \right)$  & \\
Inflation+kination  & $\sim 0$  &  1 & $\sim 0.3 \left( \displaystyle\frac{T_R}{10^7~{\rm GeV}}\right)$ &  $\sim 2 \times 10^{-17} \left( \displaystyle\frac{r}{0.01} \right)$  & \\
Particle prod. during inf.   & $- 2\epsilon$ & $ -4\epsilon (4\pi \xi - 6) (\epsilon - \eta)$  & --- &
$ \sim 2 \times 10^{-17} \left( \displaystyle\frac{r}{0.01} \right)$ & \\
2nd-order (inflation)  & $1$  & drop-off  
& $\sim 7\times 10^5  \left( \displaystyle\frac{T_{\rm reh}}{10^9~{\rm GeV}} \right)^{1/3} \left( \displaystyle\frac{M_{\rm inf}}{10^{16}~{\rm GeV}} \right)^{2/3} $ 
&$\sim 10^{-12}  \left( \displaystyle\frac{T_{\rm reh}}{10^9~{\rm GeV}} \right)^{-4/3} \left( \displaystyle\frac{M_{\rm inf}}{10^{16}~{\rm GeV}} \right)^{4/3}$   & \\
2nd-order  (PBHs)  & $2$  & drop-off   
&  $\sim 4 \times 10^{-2} \left( \displaystyle\frac{M_{\rm PBH}}{10^{20}~{\rm g}} \right)^{-1/2}$
&  $\sim 7 \times 10^{-9} \left( \displaystyle\frac{{\cal A}^2}{10^{-3}}\right)^2$ & \\
Pre-Big-Bang & $3$   & $3 - 2\mu  $  & --- & $ \sim 1.4 \times 10^{-6} \left( \displaystyle\frac{H_s}{0.15 M_{\rm pl}} \right)^4$ &\\
\hline \hline
\label{table:cosmological}
\end{tabular}
\end{center}
\end{table}
\end{landscape}
\restoregeometry

\subsection{Astrophysical sources} 
\label{subsec:astrophysical}
Here, we list astrophysical sources.  See also \cite{Schneider:2010ks}
for a collection of some astrophysical sources.  All the models we
describe in this subsection are summarized in Table
\ref{table:astrophysical}.

\bigskip
\noindent
$\bullet$ {\bf Black hole (BH) binaries and neutron star (NS) binaries}
\cite{Meacher:2015iua,Zhu:2011bd,Zhu:2012xw,TheLIGOScientific:2016wyq,Abbott:2017xzg} \vspace{2mm} \\
The GW spectra of compact binaries at low frequencies are fitted
by the power-law $\Omega_{\rm GW \ast} \propto f^{2/3}$ from the
Newtonian analysis for the inspiral phase. The cutoff is
determined by the peak frequency and given by the innermost stable
circular orbit: $f_\ast \simeq \frac{1}{6^{3/2}\pi M}\simeq 90 {\rm
  Hz}\left(\frac{M}{50 M_{\odot}}\right)^{-1}$, where $M$ is the total
mass of the binary. The parameters describing the GW background
spectrum are given by
\begin{equation}
n_{\rm GW1} = 2/3,
\qquad 
n_{\rm GW2} = {\rm cutoff},
\end{equation}

\begin{equation}
f_\ast \simeq  10^2\sim 10^3 {\rm Hz} ,
\end{equation}

\begin{equation}
\Omega_{\rm GW \ast} \simeq 10^{-9} .
\end{equation}

Note that, from the recent detection of GWs from black hole binaries
and a binary neutron star \cite{TheLIGOScientific:2017qsa}, the
amplitude of the stochastic GW background from compact binary
coalescence is estimated as $\Omega_{\rm GW}=1.8^{+2.7}_{-1.3}\times
10^{-9}$ at $25$ Hz \cite{Abbott:2017xzg}, which should be compared with 
 $\Omega_{\rm  GW}=1.1^{+1.2}_{-0.7}\times 10^{-9}$  from binary black holes alone
\cite{TheLIGOScientific:2016wyq}.  The GW background may be observed
during the next observation run (O3) of Advanced-LIGO.

\bigskip
\bigskip
\noindent
$\bullet$ {\bf White dwarf binaries} \cite{Farmer:2003pa} \vspace{2mm} \\
For the GW background from white dwarf binaries, binaries of various
masses and redshifts contribute to the background.  The resulting
slope of the spectra coming from the inspiral phase of binaries is
slightly steeper than $2/3$ for $f_{\rm crit}<f<2\times10^{-2} {\rm
  Hz}(M_{WD}/0.5 M_{\odot})$ \cite{Farmer:2003pa}, where $M_{WD}$ is
the mass of a white dwarf in a binary.  The upper cutoff of the
frequency is the one above which the inspiraling white dwarfs would
undergo Roche-lobe overflow and merge.  The critical frequency,
$f_{\rm crit}\simeq 7\times 10^{-5} {\rm Hz}(t_{\rm age}/10 {\rm
  Gyr})^{-3/8}(M_{WD}/0.5 M_{\odot})^{-5/8}$with $t_{\rm age}$ being
the age of white dwarfs, is the frequency below which the energy loss
due to GWs is not effective. For $f<f_{\rm crit}$, the slope of the GW
spectra is $10/3$ \cite{Farmer:2003pa}.  The parameters
describing the GW spectrum around the peak are thus given by
\begin{equation}
n_{\rm GW1} \simeq 2/3,
\qquad 
n_{\rm GW2} = {\rm cutoff},
\end{equation}

\begin{equation}
f_\ast \simeq  10^{-2} {\rm Hz} ,
\end{equation}

\begin{equation}
\Omega_{\rm GW \ast} \simeq 10^{-11}.
\end{equation}

\bigskip
\bigskip
\noindent
$\bullet$ {\bf Stellar core collapse (High frequency model)}
\cite{Buonanno:2004tp,Crocker:2017agi} \vspace{2mm} \\
GWs would be produced from stellar core collapse via several processes: 
the postshock convection phase, hot-bubble convection, the standing accretion 
shock instability (nonspherical  mode instability of stalled accretion shocks) 
and anisotropic neutrino emission. However, since the physics of the stellar 
core collapse is not yet fully understood, the relation of the GW 
signal to stellar progenitor properties is not well known. 
The following functional form could describe the GW spectra
predicted  in several  numerical simulations \cite{Mueller:2012sv,Ott:2012mr} 
of the stellar core collapse \cite{Crocker:2017agi}:  

\begin{equation}
\Omega_{\rm GW}(f) = \frac{8\pi Gf\xi}{3H_0^3} \int dz \frac{R_\ast(z)}{(1+z)H(z)} \left(1+\frac{f(1+z)}{a}\right)^6\exp\left(-2f(1+z)/b\right),
\end{equation}
where $\xi$ is determined by a combination of unknown parameters, such
as the mass fraction of stars undergoing core-collapse and properties of
emitted neutrinos, $a$ and $b$ (typically $5<a<150$ Hz and $10<b<400$
Hz) are free parameters of the model, $z$ is the source redshift,
$R_\ast(z)$ is the star formation rate and $H(z)$ is the Hubble
parameter.  The peak frequency may be related to the surface $g$-mode
frequency, which depends on the compactness and the surface
temperature of a massive star \cite{Mueller:2012sv}.  The spectral
shape depends on parameters. The peak frequency can vary as 
\begin{equation}
f_\ast \simeq   [10^{2},10^{3}]~ {\rm Hz} .
\end{equation}
For example, if $a=100$ and $b=200$, the
parameters spectral indices and peak frequency are
\cite{Crocker:2017agi}
\begin{equation}
n_{\rm GW1} \sim 3,
\qquad 
n_{\rm GW2} = {\rm cutoff},
\end{equation}

\begin{equation}
f_\ast \simeq 300~ {\rm Hz}.
\end{equation}
The amplitude depends on the parameter $\xi$ and can vary as \cite{Schneider:2010ks}
\begin{equation}
\Omega_{\rm GW \ast} \simeq [10^{-14},10^{-9}].
\end{equation}

\bigskip
\bigskip
\noindent
$\bullet$ {\bf Stellar core collapse (Low frequency model)} \cite{Crocker:2017agi} \vspace{2mm} \\
In some simulations of stellar core collapse
\cite{Mueller:2012sv,Ott:2012mr,Kuroda:2013rga}, the emitted GW
spectra has an additional lower peak, the origin of which may be
related to the prompt postbounce convection or the standing accretion
shock instability. The GW spectra can be fitted by the following
functional form \cite{Crocker:2017agi}:

\begin{equation}
\Omega_{\rm GW}(f) = \frac{8\pi^3 Gf A'^2}{3H_0^2} \int dz\frac{R_\ast(z)}{(1+z)H(z)} 
\exp\left(-\frac{(f(1+z)-\mu)^2}{\sigma^2}\right),
\end{equation}
where $A'$ is a scaling parameter, $\mu$ and $\sigma$ (typically
$30<\mu<200$ Hz and $10<\sigma<80$ Hz) are free parameters of the
model.  The peak frequency can vary as 
\begin{equation}
f_\ast \simeq   [10^{1},10^{2}]~  {\rm Hz} .
\end{equation}
For example, if $\mu=100$ and $\sigma=10$, the spectral
indices around the peak is \cite{Crocker:2017agi}
\begin{equation}
n_{\rm GW1} \sim 6, \qquad n_{\rm GW2} \sim 0,
\end{equation}

\begin{equation}
f_\ast \simeq 40~ {\rm Hz}.
\end{equation}
The amplitude depends on the parameter $A'$ and varies as \cite{Schneider:2010ks}
\begin{equation}
\Omega_{\rm GW \ast} \simeq [10^{-14},10^{-9}].
\end{equation}

\bigskip
\bigskip
\noindent
$\bullet$ {\bf r-mode instability of NSs} \cite{Ferrari:1998jf,Zhu:2011pt}  \vspace{2mm} \\
Rapidly rotating neutron stars suffer from the so-called r-mode
instability, the instability of toroidal perturbations by the emission
of GWs \cite{Andersson:1997xt, Friedman:1997uh}.  The rotational
energy is converted into GWs and hence the maximum frequency of the
gravitational radiation is determined by the initial rotational
frequency of a neutron star which is approximately limited by the Kepler frequency \cite{Friedman:1989zzb}.
The parameters describing the GW background spectrum are given by
\begin{equation}
n_{\rm GW1} \simeq 2,
\qquad 
n_{\rm GW2} = {\rm cutoff},
\end{equation}

\begin{equation}
f_\ast \simeq 1.5\times 10^3 {\rm Hz}\sqrt{\frac{M}{M_{\odot}}\left(\frac{10{\rm km}}{R}\right)^3}, 
\end{equation}

\begin{equation}
\Omega_{\rm GW \ast} \simeq [10^{-12},10^{-8}],
\end{equation}
where $M$ and $R$ are the mass and the radius of a neutron star.

\bigskip
\bigskip
\noindent
$\bullet$ {\bf Magnetar} \cite{Regimbau:2005ey,Marassi:2010wj} \vspace{2mm} \\
Magnetars are neutron stars with extremely large magnetic fields
($>10^{14}$G).  These large magnetic fields deform the shape of
neutron stars and cause the emission of significant GWs if these stars
are rapidly rotating and the magnetic dipole axis is different from
the rotation axis \cite{Regimbau:2005ey,Marassi:2010wj}.  The
parameters characterizing the GW background spectrum are given by
\begin{equation}
n_{\rm GW1} \simeq 3,
\qquad 
n_{\rm GW2} = {\rm cutoff},
\end{equation}

\begin{equation}
f_\ast \simeq 10^3 {\rm Hz} ,
\end{equation}

\begin{equation}
\Omega_{\rm GW \ast} \simeq 10^{-16}\sim 10^{-8}.
\end{equation}

\bigskip
\bigskip
\noindent
$\bullet$ {\bf Superradiant instabilities} \cite{Brito:2017wnc,Yoshino:2013ofa} \vspace{2mm} \\
Light scalar fields around spinning black holes can induce
superradiant instabilities which transfer the rotational energy of the
black holes to trigger the growth of a bosonic condensate outside the
horizon.  Although superradiant instabilities produces ``holes'' in
the BH mass/spin plane (``Regge plane'') determined by the
measurements of GWs from resolvable BH sources, a population of
massive BH-bosonic condensates can form a stochastic background of GW
from the condensate.  The emitted GWs are nearly monochromatic with
the frequency $\sim m_s/\pi$, where $m_s$ is the mass of the scalar
field.  The spectrum depends on the formation rate and the number
density of BHs and strongly on the spin distribution of the BHs.
According to \cite{Brito:2017wnc}, the parameters characterizing the
GW background spectrum are roughly given by

\begin{equation}
n_{\rm GW1} = 1\sim 7,
\qquad 
n_{\rm GW2} < 0,
\end{equation}

\begin{equation}
f_\ast \simeq \frac{m_s}{\pi}\simeq 5\times 10^2 \left(\frac{m_s}{10^{-12}{\rm eV}}\right) {\rm Hz} ,
\end{equation}

\begin{equation}
\Omega_{\rm GW \ast} \siml 10^{-6}.
\end{equation}

\bigskip

\begin{table}
\begin{center}
  \caption{Astrophysical GW sources }
\label{table:astrophysical}
\vspace{2mm}
\begin{tabular}{r|ccccl}
\hline \hline
source & $n_{\rm GW1}$ & $n_{\rm GW2}$ & $f_\ast$ [Hz] & $\Omega_{\rm GW}$ &  \\ \hline 
Neutron star merger & $2/3$  & cutoff & $\sim 10^3$  &  $\sim 10^{-9}$ &  \\
Black hole merger  & $2/3$  & cutoff &  $\sim 10^2$  & $\sim 10^{-9}$ &  \\
White dwarf  & $2/3$ & cutoff & $\sim 10^{-2}$ & $\sim 10^{-11}$ & \\
Stellar core collapse I (High frequency model) &  ---  & ---  & $[10^2, 10^3]$ & $[10^{-14},10^{-9}]$   & \\
Stellar core collapse II (Low frequency model)  & ---  & ---   & $[10^1, 10^2]$ & $[10^{-14},10^{-9}]$ &  \\
Neutron star r-mode & $2$  & cutoff  & $\sim 10^{3}$ &  $[10^{-12},10^{-8}]$ &  \\
Magnetar    & $3$ & cutoff & $\sim 10^3$ & $[10^{-16}, 10^{-8}]$ & \\
Superradiant instabilities    &  $1\sim 7 $ & $<0$ & --- & $< 10^{-6}$ & \\
\hline \hline
\end{tabular}
\end{center}
\end{table}


\section{Methodology}
\label{sec:methodology}

As mentioned in the introduction, the main purpose of this paper is to
investigate to what extent we can probe the source of the stochastic
GW background with future GW experiments by looking at the spectral
shapes, more specifically, the spectral indices of the GW power
spectrum.  To pursue this, we adopt the Fisher matrix analysis to
study expected constraints from future GW observations on the
parameters characterizing the GW spectrum such as the amplitude and
the spectral indices. In Section \ref{sec:fisher}, we first summarize
the formalism of the Fisher matrix analysis.  Then in Section
\ref{sec:parametrization}, we describe how to parametrize the GW
spectrum.

\subsection{Fisher analysis}
\label{sec:fisher}
Here, we briefly describe the statistics related to the detection of
the stochastic background \cite{Allen:1997ad} and Fisher matrix
formalism \cite{Seto:2005qy} which is used to forecast constraints on
parameters describing the GW spectrum.

Let us decompose the metric perturbation $h_{ij}$ into its Fourier
modes $\tilde{h}_\lambda$ and denote the two independent polarization
states as
\begin{equation}
  h_{ij}(t,\textbf{x})=\sum_{\lambda=+,\times}^{}\int df\int_{S^2}d{\bf\Omega}
  \tilde{h}_\lambda(f,{\bf\Omega}) \epsilon_{ij}^{\lambda}
({\bf\Omega})e^{i2\pi f(t-{\bf\Omega}\cdot\textbf{x})}.
\end{equation}
where ${\bf \Omega} = \cos \phi \, \sin \theta \, {\bm e}_x + \sin
\phi \, \sin \theta \, {\bm e}_y + \cos \theta \, {\bm e}_z$ is a
vector pointing to a direction on the two-sphere specified by the standard
polar and azimuthal angle $\phi$ and $\theta$.  The polarization
tensors $\epsilon_{ij}^\lambda$, where $\lambda$ indicates the plus
($+$) and cross ($\times$) polarization, satisfy the symmetric and
transverse-traceless conditions and are normalized as
$\sum_{i,j}^{}\epsilon_{ij}^{\lambda}(\epsilon_{ij}^{\lambda^{\prime}})^*=2\delta^{\lambda\lambda^{\prime}}$.

The stochastic GW search is performed by taking a cross correlation of
signals between two detectors. Let us label two different detectors by
$I$ and $J$.  Then the cross correlated signal is given by
\begin{equation}
S=\int^{T/2}_{-T/2}dt\int^{T/2}_{-T/2}dt^{\prime} s_I (t)s_J (t^{\prime})Q (t,t^{\prime}),
\end{equation}
where $T$ is the observation time, $Q (t,t^{\prime})$ is a filter
function, and $s_I (t)=h_I(t)+n_I(t)$ is the output signal of the
detector $I$ composed of the GW signal $h_I(t)$ and detector noise
$n_I(t)$. Since noises between different detectors have no
correlation, $\langle s_I (t)s_J (t^{\prime})\rangle\simeq \langle h_I
(t)h_J (t^{\prime})\rangle$, the mean value of the signal can be
expressed in the Fourier space as
\begin{equation}
\mu\equiv\langle S \rangle =\int^{\infty}_{-\infty}df\int^{\infty}_{-\infty}df^{\prime}\delta_T (f-f^{\prime})\langle\tilde{h}^*_I (f)\tilde{h}_J (f^{\prime})\rangle\tilde{Q} (f^{\prime}),
\end{equation}
where the tilde denotes Fourier-transformed quantities,
$\langle\cdots\rangle$ denotes the ensemble average, and $\delta_T
(f)\equiv\int^{T/2}_{-T/2}dt e^{-2\pi ift}$.  The GW signal $\tilde{h}_I(f)$ is described by using $F^\lambda_I$
which describes the response of the detector as
\begin{equation}
\tilde{h}_I (f)=\sum_\lambda\int d\hat{\bf \Omega} \tilde{h}_\lambda (f,{\bf \Omega})
e^{-2\pi if\hat{\bf \Omega}\cdot{\bf x}_I} F^\lambda_I (f,{\bf \Omega}),
\end{equation}
where ${\bf x}_I$ is the position of the detector.
Using the relation of
\begin{equation}
\langle \tilde{h}^*_\lambda (f,{\bf \Omega})
\tilde{h}_{\lambda^{\prime}} (f^{\prime},{\bf \Omega^{\prime}})\rangle
=\frac{3H_0^2}{32\pi^3}\delta^{(2)} ({\bf \Omega,\Omega^{\prime}})
\frac{1}{2}\delta_{\lambda\lambda^{\prime}}\delta (f-f^{\prime})
|f|^{-3}\Omega_{\rm GW} (|f|),
\end{equation}
with $ \delta^{(2)} ({\bf \Omega,\Omega^{\prime}}) = \delta (\phi - \phi') \, \delta (\cos \theta - \cos \theta')$, 
the cross correlation signal is given by
\begin{equation}
\mu=\frac{3H_0^2}{20\pi^2} T
\int^{\infty}_{-\infty}df|f|^{-3} \gamma_{IJ} (f)
\Omega_{\rm GW} (f)\tilde{Q} (f),
\label{Eqmu}
\end{equation}
where the overlap reduction function $\gamma_{IJ}$ is given by
detector responses as \cite{Nishizawa:2009bf}
\begin{equation}
\gamma_{IJ} (f)\equiv\frac{5}{8\pi}\int d\hat{\bf \Omega}
 (F^+_IF^+_J+F^{\times}_IF^{\times}_J)
e^{-2\pi if\hat{\bf \Omega}\cdot ({\bf x}_I-{\bf x}_J)}\,.
\end{equation}
In the weak-signal assumption, the variance of the correlation signal
is
\begin{eqnarray}
  \sigma^2
  &\equiv&\langle S^2\rangle-\langle S \rangle^2\approx\langle S^2\rangle
\nonumber\\
&=&\int^{T/2}_{-T/2}dt\int^{T/2}_{-T/2}dt^{\prime} \langle s_I (t)s_J (t)s_I (t^{\prime})s_J (t^{\prime})\rangle Q (t) Q (t^{\prime})\nonumber\\
&\approx&\frac{T}{4}
\int^{\infty}_{-\infty}df S_{n,I} (|f|)S_{n,J} (|f|)|\tilde{Q} (f)|^2\,.
\label{Eqsigma}
\end{eqnarray}
In the last step, we used $\langle s_I (t)s_J (t)s_I (t^{\prime})s_J
(t^{\prime})\rangle\simeq \langle n_I (t)n_I
(t^{\prime})\rangle\langle n_J (t)n_J (t^{\prime})\rangle$ and
transformed the equation into Fourier space. The noise spectral
density $S_{n,I} (f)$ is defined by $\langle n_I (f)^*n_I (f^{\prime})\rangle\equiv S_{n,I}
(f)\delta (f-f^\prime)/2$.  Then we find that the signal-to-noise
ratio (SNR) $\rho\equiv\mu/\sigma$ is maximized by choosing the optimal
function as $\tilde{Q} (f)\propto \frac{\gamma_{IJ} (|f|)\Omega_{\rm
    GW} (|f|)}{|f|^3S_{n,I} (|f|)S_{n,J} (|f|)}$, and can be written
as
\begin{equation}
\rho_{IJ}=\frac{3H_0^2}{10\pi^2} \sqrt{2T}
\left[\int^{\infty}_{0}df\frac{|\gamma_{IJ} (f)|^2
\Omega_{\rm GW} (f)^2}{f^6S_{n,I} (f)S_{n,J} (f)}\right]^{1/2}.
\label{rhoIJ}
\end{equation}
For a network of $N$ detectors, SNR is 
\begin{equation}
\rho=
\left[\sum^N_{I=1}\sum^N_{J<I}\rho_{IJ}^2\right]^{1/2}.
\label{rho}
\end{equation}
Describing Eqs. (\ref{Eqmu}) and (\ref{Eqsigma}) in terms of the
discrete Fourier transform, the signal and its variance are rewritten
as
\begin{eqnarray}
\langle \mu\rangle =2\sum_i\frac{3H_0^2}{20\pi^2}\frac{\delta f_i}{\Delta f}f_i^{-3}\gamma_{IJ}(f_i)\Omega_{\rm GW}(f_i)\tilde{Q}(f_i)
\equiv\sum_i\langle \mu_i\rangle ,
\label{signal_i}
\end{eqnarray}
\begin{equation}
\sigma^2=2\sum_i\frac{1}{4}\frac{\delta f_i}{\Delta f}S_{n,I}(f_i) S_{n,J}(f_i)|\tilde{Q}(f_i)|^2\equiv\sum_i\sigma_i^2,
\label{noise_i}
\end{equation}
where $i$ labels each frequency bin with center frequency $f_i$
and width $\delta f_i$, which is taken to be much larger than $\Delta
f\equiv T^{-1}$.

Let us assume that the data $\mu_i$ have a Gaussian distribution
around the mean value $\langle \mu_i\rangle$, and then the likelihood
function ${\cal L}$ is defined by the product of all the probabilities
of frequency bins as
\begin{equation}
{\cal L}=\prod_i \frac{1}{\sqrt{2\pi\sigma_i^2}}\exp\left[-\frac{(\mu_i-\langle \mu_i\rangle )^2}{2\sigma_i^2}\right].
\label{likelihood}
\end{equation}
The mean value can be replaced by the theoretically expected value
$\mu(f_i;\hat\theta^{\rm fid})$, where $\hat\theta^{\rm fid}$ denotes
the fiducial values of model parameters when we investigate expected
constraints for model parameters from future observations.  Maximizing
the likelihood function is equivalent to minimizing $\delta\chi^2$,
which is defined by
\begin{eqnarray}
\label{eq:chi2}
\delta\chi^2(\hat\theta;\hat\theta^{\rm fid})
&=&-2\ln \left( \frac{{\cal L}(\hat\theta;{\hat\theta^{\rm fid}})}{{\cal L}(\hat\theta^{\rm fid};{\hat\theta^{\rm fid}})}  \right)\nonumber\\
&=&\left (\frac{3H_0^2}{10\pi^2}\right)^2 4T_{\rm obs}
\int^{\infty}_{0}df\frac{|\gamma_{IJ} (f)|^2[\Omega_{\rm GW} (f;\hat\theta)-\Omega_{\rm GW} (f;\hat\theta^{\rm fid})]^2}{f^6S_{n,I} (f)S_{n,J} (f)},
\end{eqnarray}
where $\hat\theta$ are the parameter values assumed to fit the
data $\mu_i$.  In the second step, we have substituted
Eq.~(\ref{likelihood}), and used $\langle \mu_i\rangle$ and $\sigma_i$
in Eqs. (\ref{signal_i}) and (\ref{noise_i}).  Assuming that the
likelihood function can be approximated by a Gaussian distribution
around the maximum in the parameter space, we can expand
$\delta\chi^2$ as
\begin{equation}
 \delta\chi^2(\hat\theta;\hat\theta^{\rm fid}) =   \sum_{lm}(\theta_l-\theta_l^{\rm fid})  {\cal F}_{lm}(\hat\theta^{\rm fid}) (\theta_m-\theta_m^{\rm fid})\,,
 \label{eq:chi2_Fisher}
\end{equation}
where $l,m$ run over model parameters.  The Fisher information matrix
describes the local curvature of the likelihood function ${\cal L}$
and is defined as
\begin{equation}
{\cal F}_{lm}(\hat\theta^{\rm fid})\equiv -\frac{\partial^2\ln{\cal L}}{\partial\theta_l\partial\theta_m}\,.
\label{Fisher}
\end{equation} 
Then the expected error in the parameter $\theta_l$ is given by
$\sigma_{\theta_l}=\sqrt{ ({\cal F}^{-1})_{ll}}$.  Substituting the
likelihood into Eq.~(\ref{Fisher}), and using $\langle \mu_i\rangle$
and $\sigma_i$ in Eqs. (\ref{signal_i}) and (\ref{noise_i}), the
Fisher matrix is given by \cite{Seto:2005qy}
\begin{equation}
{\cal F}_{lm,IJ}=\left (\frac{3H_0^2}{10\pi^2}\right)^2 2T_{\rm obs}
\int^{\infty}_{0}df\frac{|\gamma_{IJ} (f)|^2\partial_{\theta_l}
\Omega_{\rm GW} (f)\partial_{\theta_m}\Omega_{\rm
GW} (f)}{f^6S_{n,I} (f)S_{n,J} (f)}\,.
\end{equation}
For multiple detectors, the Fisher matrix can be written as
\begin{equation}
{\cal F}_{lm}=\left (\frac{3H_0^2}{10\pi^2}\right)^2 2T_{\rm obs}\sum^N_{I=1}\sum^N_{J<I}
\int^{\infty}_{0}df\frac{|\gamma_{IJ} (f)|^2\partial_{\theta_l}
\Omega_{\rm GW} (f)\partial_{\theta_m}\Omega_{\rm
  GW} (f)}{f^6S_{n,I} (f)S_{n,J} (f)}\,.
\label{Eq:Fisher}
\end{equation}
Note that this expression is obtained by using a weak-signal limit
$h_I\ll n_I$ in Eq.~\eqref{Eqsigma}, which cannot be used when the SNR
is large.  The authors of \cite{Kudoh:2005as} have found that
overestimation of the SNR occurs when this approximation breaks down
and the effect becomes significant above SNR $\sim 100$.  Therefore,
the Fisher analysis based on Eq.~\eqref{Eq:Fisher} would not give good
estimations for SNR $\gtrsim 100$.

\begin{figure}
  \begin{center}
    \includegraphics[width=6.4in]{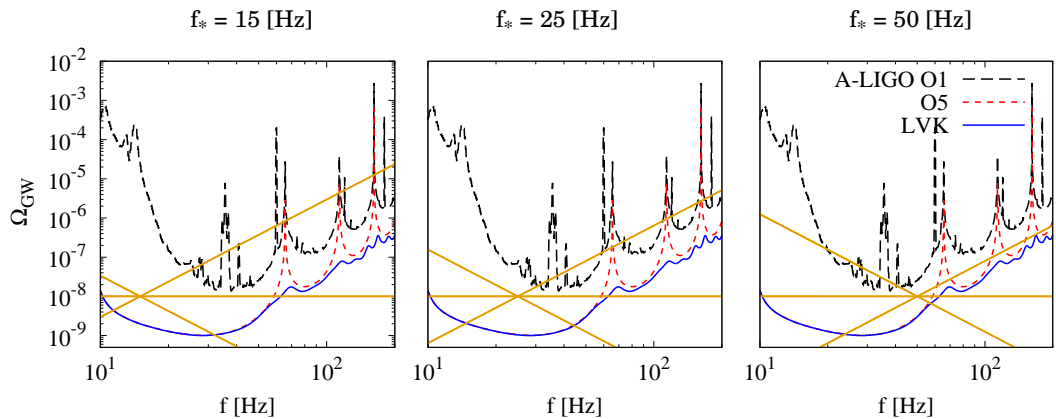}
  \end{center}
  \vspace{-10mm}
  \caption{Comparison of the sensitivity curve and the GW spectra
    given in Eq.~\eqref{eq:GW_spectrum} for the cases with $f_\ast =
    15~{\rm Hz}$ (left), $25~{\rm Hz}$ (middle) and $50~{\rm Hz}$
    (right).  We show sensitivities of aLIGO O1 (black dashed), O5
    (red dotted), and 4 detector network by aLIGO-aVirgo-KAGRA (LVK,
    blue solid).  For the GW spectra (yellow solid), we fix the
    fiducial value of $\Omega_\ast$ at $10^{-8}$ and $n_{\rm GW1}$ and
    $n_{\rm GW2}$ take values on $-3, 0, 3$.  }
  \label{fig:sensitivity}
\end{figure}

In the following analysis, we assume $3$-year observations of the
world-wide four-detector network consisting of LIGO-Hanford and
LIGO-Livingstone with the O5 sensitivity and Advanced-Virgo and KAGRA with
their design sensitivities.  In Fig.~\ref{fig:sensitivity}, the
sensitivity curves are shown for aLIGO O1, O5 and the four-detector
network aLIGO-aVirgo-KAGRA.  The sensitivity curve represents the
threshold of SNR $=1$ in the frequency range $[f,f+\Delta f]$, where
we take $\Delta f=f/10$ \cite{Kudoh:2005as}.  We use the noise spectra
shown in \cite{Aasi:2013wya}, whose data is provided by the LIGO
document control center \cite{ref:sensitivity}.  The overlap reduction
function is calculated following \cite{Nishizawa:2009bf}.  For the
frequency integration, we take the low frequency cutoff at $10$~Hz and
high frequency cutoff at $200$~Hz. 

We also provide results for DECIGO, whose sensitivity and overlap
reduction function can be found in \cite{Kuroyanagi:2010mm} and
\cite{Kudoh:2005as}, respectively.  The sensitivity curve for DECIGO
is shown in Fig. \ref{fig:spectrum_decigo} in the next section.  For
the analysis, we assume $3$-year observation and the low frequency
cutoff is taken at $10^{-3}$~Hz and high frequency cutoff is taken at
$100$~Hz.

\subsection{Parameterizing the GW spectrum}
\label{sec:parametrization}

Although, as we discussed in Section~\ref{sec:GW_source}, most sources
of stochastic GW background have a broken power-law shape, a single
observation may only be able to see a limited frequency range of the
spectrum and may not cover the typical frequency at which the GW
spectrum changes its scale dependence.  In this case, the single
power-law fit would be sufficient.  Therefore, we make two types
of analysis where we parametrize the GW spectrum as follows:

\bigskip
\noindent
(i) Single power-law 
\begin{equation}
\label{eq:single_power}
\Omega_{\rm GW} (f)  = \Omega_{\rm GW \ast} \left( \displaystyle\frac{f}{f_\ast} \right)^{n_{\rm GW}}\,.
\end{equation}

\bigskip
\noindent
(ii) Broken power-law 
\begin{equation}
\label{eq:broken_power}
\Omega_{\rm GW} (f)  =
\begin{cases}
      &  \Omega_{\rm GW \ast} \left( \displaystyle\frac{f}{f_\ast} \right)^{n_{\rm GW1}}  \qquad \text{for}~~ f < f_\ast, \\ \\
      & \Omega_{\rm GW \ast} \left( \displaystyle\frac{f}{f_\ast} \right)^{n_{\rm GW2}}  \qquad \text{for}~~ f > f_\ast, 
\end{cases}
\end{equation}
where $\Omega_{\rm GW \ast}$ is the amplitude at the reference frequency $f_\ast$.

\begin{figure}
  \begin{center}
    \begin{tabular}{c}
      \begin{minipage}{0.4\hsize}
        \begin{center}
          \includegraphics[width=2.5in]{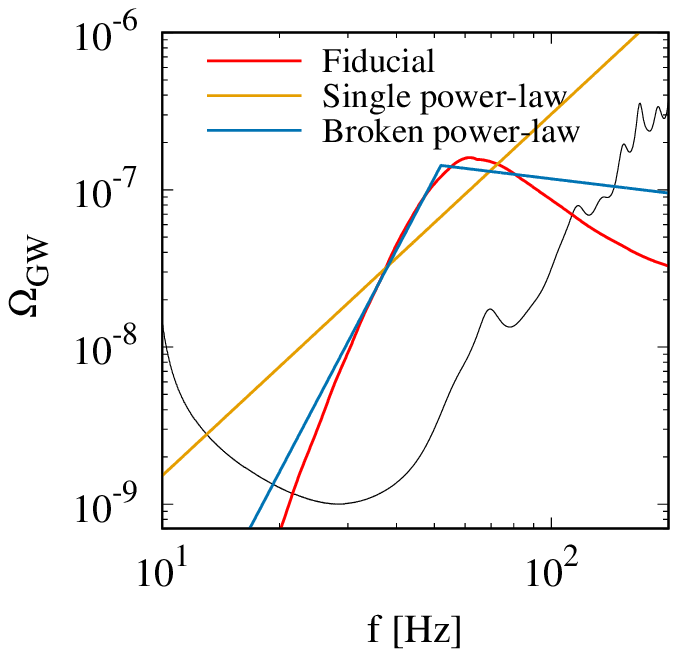}
        \end{center}
      \end{minipage}
      
      \begin{minipage}{0.4\hsize}
        \begin{tabular}{|l|cc|}
          \hline
          ~ & SNR & $\delta\chi^2$ \\
          \hline
          Single power-law & 70.7 & 1440 \\
          Broken power-law & 80.0 & 47.4 \\
          Fiducial & 80.3 & 0 \\
          \hline
        \end{tabular}
      \end{minipage}
    \end{tabular}
    \caption{Left panel: Example of single and broken power-law
      fittings.  The red line is the GW background generated by
      superradiant instabilities (taken from Fig. 2 of
      \cite{Brito:2017wnc}).  The yellow and blue lines are the
      best-fit spectra for single and broken power-law fits
      respectively. The black line shows the sensitivity curve of
      aLIGO-aVirgo-KAGRA. Right panel: SNR and $\delta\chi^2$ for the
      best-fit spectra, calculated assuming the aLIGO-aVirgo-KAGRA
      sensitivity.  }
       \label{fig:superradiant}
  \end{center}
\end{figure}

\begin{figure}
  \begin{center}
    \includegraphics[width=4.9in]{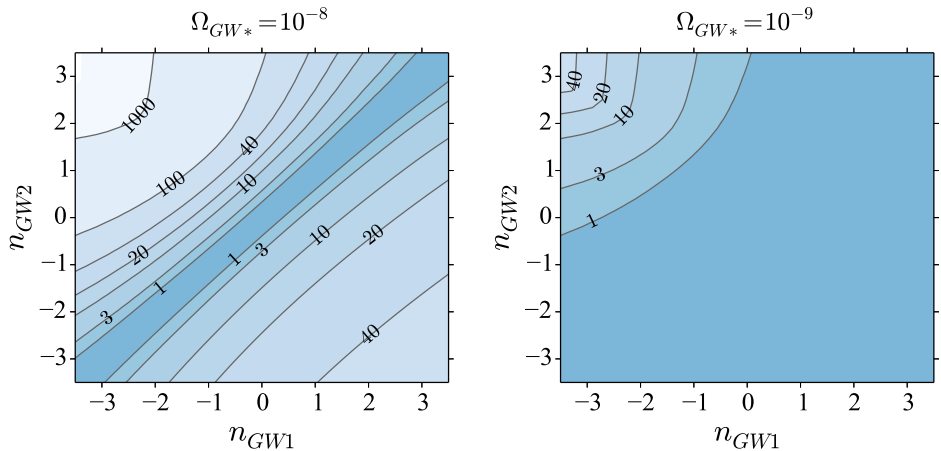}
  \end{center}
  \vspace{-5mm}
  \caption{Contour plot of $\delta\chi^2|_{\rm diff}$ in the $n_{\rm
      GW1}$ -- $n_{\rm GW2}$ plane, showing the difference of
    $\delta\chi^2$ between the single power-law fit,
    Eq.~\eqref{eq:single_power}, and broken power-law fit,
    Eq.~\eqref{eq:broken_power}.  The pivot frequency is fixed at
    $f_\ast = 25$~Hz and the amplitude is taken as $\Omega_{\rm GW
      \ast} = 10^{-8}$ in the left panel and $10^{-9}$ in the right
    panel.  The specification of aLIGO-aVirgo-KAGRA is assumed.}
 \label{fig:SN_Omega_n_single:diff}
\end{figure}

Here we demonstrate how the choice of the template spectrum affects
the value of $\delta \chi^2$, which describes the goodness of the
fit. As an example, we take the case of GWs from superradiant
instabilities (the most pessimistic case for $m_s=10^{-12.5}{\rm eV}$)
\cite{Brito:2017wnc}\footnote{
Note that this model is already ruled
out since the SNR of the predicted spectrum is SNR=6.64 for aLIGO O1
with a single power-law template (SNR=7.25 with the template of
 fiducial spectrum).
 }.  In Fig.~\ref{fig:superradiant}, the GW
spectrum from the model, and the best-fit spectra for single and
broken power-law templates are shown.  In the right panel of
Fig.~\ref{fig:superradiant}, we tabulate the SNR expected for
aLIGO-aVirgo-KAGRA and $\delta \chi^2$ for each case.  We see that,
when the fitting is performed with broken power-law templates, the
value of SNR improves more than $10\%$ compared with the case fitted
by the single power-law template. We also find the value of $\delta
\chi^2$ significantly differs between single and broken power-law fits.

Let us extend the discussion to general cases with different values
of the spectral indices $n_{\rm GW1}$ and $n_{\rm GW2}$.  In
Fig.~\ref{fig:SN_Omega_n_single:diff}, we show
the contour plot of $\delta \chi^2|_{\rm diff}$ whose definition is
\begin{equation} 
 \delta\chi^2|_{\rm diff} \equiv \delta \chi^2 |_\textrm{single power-law} - \delta \chi^2 |_\textrm{broken power-law},
\end{equation}
where $ \delta \chi^2 |_\textrm{single power-law}$ is calculated
assuming that the fiducial spectrum is the broken-power spectrum with
$n_{\rm GW1}, n_{\rm GW2}$ and $f_\ast = 25$~Hz and searching the
best-fit spectrum using single power-law templates, while $\delta
\chi^2 |_\textrm{broken power-law}$ is calculated by fitting with
broken power-law templates.  Thus $\delta\chi^2|_{\rm diff}$ describes
how much the fit gets worse when we use single power-law templates for
broken power-law fiducial spectrum.  Notice that, by definition,
$\delta \chi^2 |_\textrm{broken power-law best-fit}=0$.  We show two
cases where the amplitude is assumed as $\Omega_{\rm GW\ast} =10^{-8}$
and $10^{-9}$, and $\delta \chi^2$ is calculated by assuming the
sensitivity of aLIGO-aVirgo-KAGRA.  Note that the case with $n_{\rm
  GW1} = n_{\rm GW2}$ reduces to the case of a single power-law
template, so $\delta \chi^2|_{\rm diff}$ is zero for $n_{\rm GW1} =
n_{\rm GW2}$.  We find that $\delta \chi^2|_{\rm diff}$ increases when
the fiducial model deviates more from a single power-law case, i.e.,
as the broken power-law nature becomes more evident.  This shows that
we should use a broken power-law form for templates when the actual
model has a break in the spectrum inside the frequency range to which
the observation is sensitive.  Therefore, we suggest that both
single and broken power-law templates should be investigated when one
analyzes a stochastic GW spectrum.  It is also worth mentioning that
$\delta \chi^2|_{\rm diff}$ increases when SNR is larger.  As seen in
the figure, $\delta \chi^2|_{\rm diff}$ for the case with a negative
$n_{\rm GW1}$ and a positive $n_{\rm GW2}$ is larger than that for a
positive $n_{\rm GW1}$ and a negative $n_{\rm GW2}$, since the former
has a spectrum with a downward convex shape, which is detected with
larger SNR for the same fiducial amplitude $\Omega_{\rm GW\ast}$.
This tendency can be also seen by comparing the cases of different
fiducial amplitude, $\Omega_{\rm GW\ast} =10^{-8}$ and $\Omega_{\rm
  GW\ast} =10^{-9}$.

\section{Expected constraints from aLIGO-aVirgo-KAGRA and DECIGO}
\label{sec:result}

As we have seen in Sec. \ref{sec:GW_source}, stochastic backgrounds
have different power-law dependence at high and low frequencies and
the single power-law fit is not always a good approximation.  We have
introduced the broken power-law as the next step after the single
power-law and, in Sec. \ref{sec:methodology}, we demonstrated that the
broken power-law template improves the fit.  Note that, in some
models, change of the spectral dependence is not sharp at $f_*$ and
the broken power-law may not be the best choice of template.  However,
preparing precise templates is challenging for some models and, on top
of that, we lose generality if we assume a specific model.  Therefore
here we investigate only the single and broken power-law cases as a
simple setup.

Now we discuss to what extent we can probe the origin of GWs by
looking at the shapes of the GW spectrum, which is characterized by
the amplitude at the reference scale $f_\ast$ (this frequency also
corresponds to the break of the power law for the broken power-law
case) and spectral indices ($n_{\rm GW}$ for the single-power case,
$n_{\rm GW1}$ and $n_{\rm GW2}$ for the broken power-law case).
First, we discuss the analysis using the single power-law templates
with aLIGO-aVirgo-KAGRA sensitivity, then the case of the broken
power-law ones follows. Finally, we also present the results for
DECIGO in Section \ref{sec:DECIGO}.

\subsection{Single power-law case}
\label{subsec:single}
\begin{figure}
  \begin{center}
    \includegraphics[width=3.4in]{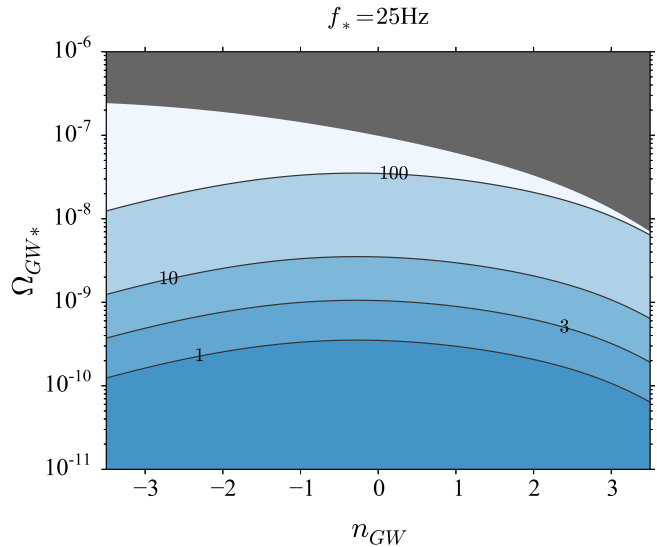}
  \end{center}
  \vspace{-5mm}
  \caption{Contour plot representing the SNR in the $\Omega_{\rm
      GW\ast}$--$n_{\rm GW}$ plane for the case where the fiducial
    spectrum has single power-law shape.  The reference frequency is
    taken to be $f_\ast=25$~Hz.  The gray color represents the
    parameter space which is already constrained by aLIGO O1
    at 2$\sigma$ level \cite{TheLIGOScientific:2016dpb}.
  }
  \label{fig:SN_Omega_n_single}
\end{figure}

The single power-law templates have two free parameters to be
determined: the amplitude $\Omega_{\rm GW\ast}$ (at the reference
scale $f_\ast$) and the spectral index $n_{\rm GW}$. Let us first show
the parameter space which will be accessible with the future
experiment sensitivity.  In Fig.~\ref{fig:SN_Omega_n_single}, we show
the expected SNR from aLIGO-aVirgo-KAGRA in the $\Omega_{\rm
  GW\ast}$--$n_{\rm GW}$ plane. Here the reference frequency is taken
to be $f_\ast = 25$ Hz, at which aLIGO O5 is most sensitive. The gray
region in the figure is already excluded by aLIGO O1
~\cite{TheLIGOScientific:2016dpb} at 2$\sigma$ level. Note
that this prediction changes depending on the fiducial frequency
$f_\ast$.  Since the sensitivity curve is not symmetric around $f_\ast
= 25$ Hz as seen in Fig.~\ref{fig:sensitivity}, contours of SNR are
also slightly asymmetric.  Also, since aLIGO O1 is most
sensitive around $f \simeq 40$ Hz, a bluer spectrum tends to be
excluded when we take $f_\ast = 25$ Hz.

\begin{figure}
  \begin{center}
    \includegraphics[width=5.4in]{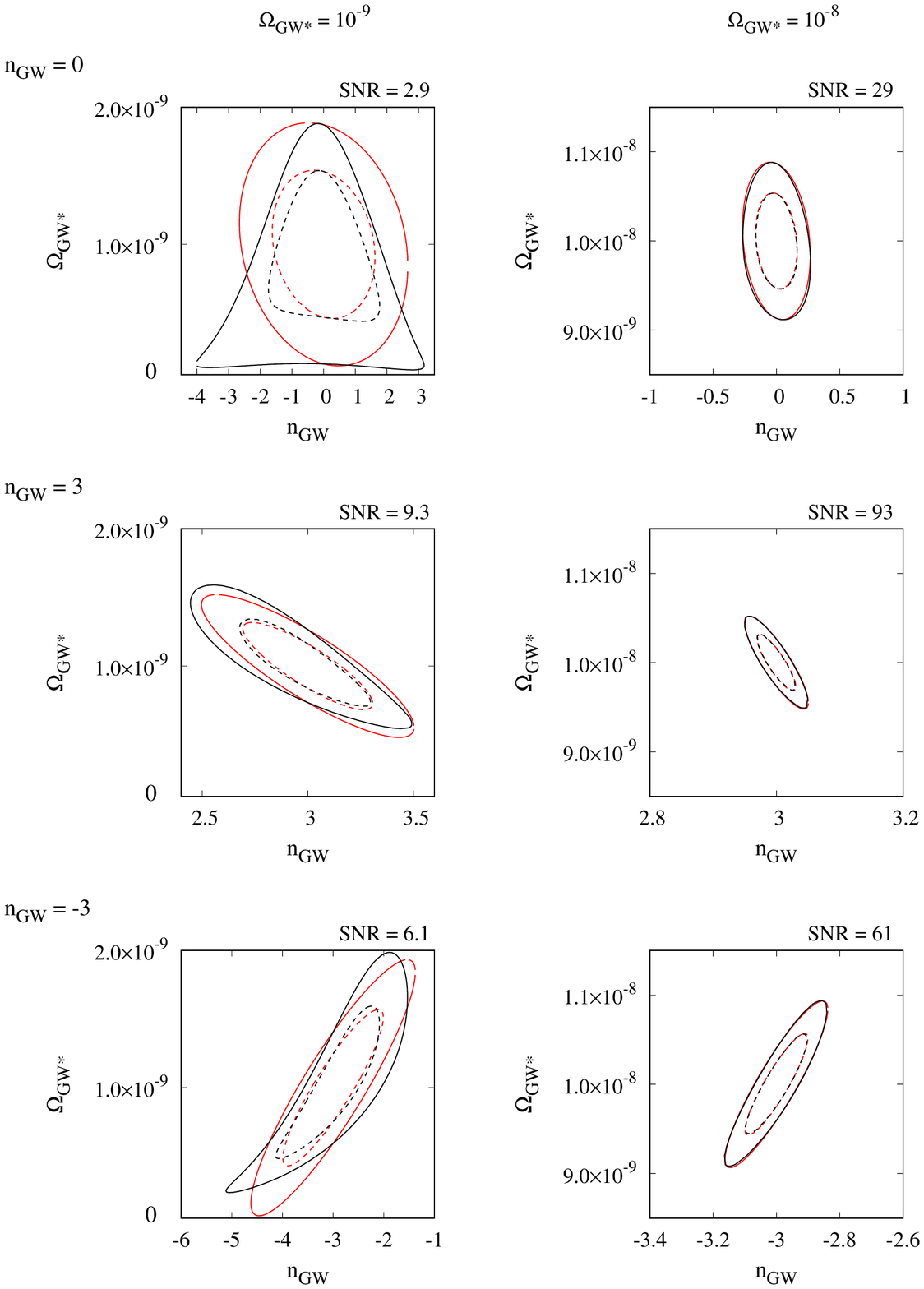}
  \end{center}
  \vspace{-10mm}
  \caption{Expected constraints on parameters $\Omega_{\rm GW\ast}$
    and $n_{\rm GW}$ for a single power-law case. We show 1$\sigma$
    (dashed line) and 2$\sigma$ (solid line) limits estimated using
    the Fisher matrix analysis (red) and the $\chi^2$ analysis
    (black).  Each panel corresponds to different fiducial values,
    $\Omega_{\rm GW\ast}=10^{-9}$ (left), $10^{-8}$ (right) and
    $n_{\rm GW}=0$ (top), $3$ (middle), $-3$ (bottom).  The reference
    frequency is taken to be $f_\ast=25$~Hz. }
  \label{fig:const_Omega_nGW_single}
\end{figure}

\begin{figure}
  \begin{center}
    \includegraphics[width=3.0in]{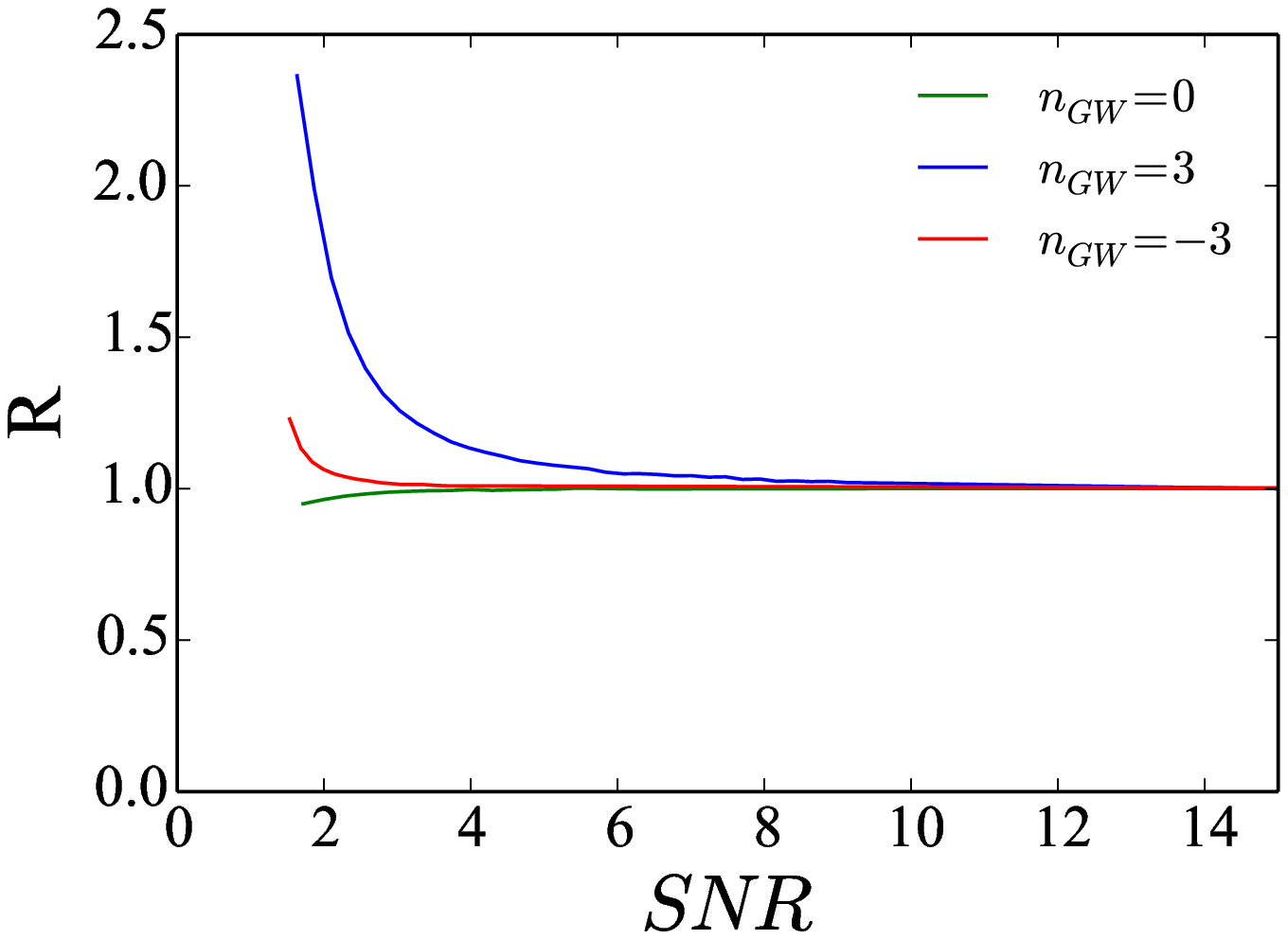}
  \end{center}
  \vspace{-5mm}
  \caption{Comparison of the area of $1\sigma$ allowed region in the
    $\Omega_{\rm GW\ast}$--$n_{\rm GW}$ parameter space.  The ratio $R$ is
    plotted as a function of SNR. Here the region of $\Omega_{\rm
      GW\ast}<0$ in the Fisher prediction is not included as the
    area. }
  \label{fig:area_single_power}
\end{figure}

Once we detect a GW background, we would be able to perform a
parameter estimation and obtain the values of $\Omega_{\rm GW\ast}$
and $n_{\rm GW}$ with error bars.  In
Fig.~\ref{fig:const_Omega_nGW_single}, we demonstrate some examples of
parameter constraints for different fiducial values of $\Omega_{\rm
  GW\ast}$ and $n_{\rm GW}$ assuming the aLIGO-aVirgo-KAGRA
observation.  See \cite{Seto:2005qy}, for the first attempt to
estimate errors on $\Omega_{\rm GW\ast}$ and $n_{\rm GW}$ for the
analysis with a single power-law.  We show two different contours;
black curves represent results from the $\chi^2$ analysis with
constant $\delta\chi^2$ slices at $\delta\chi^2=2.3$ ($1\sigma$) and
$6.18$ ($2\sigma$), while red curves are those obtained by calculating
the Fisher matrix under the assumption of a Gaussian likelihood shape
around the reference parameter value.  From the figure, we find that
the shape of the likelihood function deviates from Gaussian when the
SNR is low (left column, $\Omega_{\rm GW\ast}=10^{-9}$), while the
Fisher prediction shows good agreement with the contours from $\chi^2$
analysis when the SNR is high (right column, $\Omega_{\rm
  GW\ast}=10^{-8}$).  Therefore, by comparing the areas of the
1$\sigma$ allowed parameter space we may judge whether the Fisher
matrix provides good estimate for predicting future constraints.  Here
we define the following quantity:
\begin{equation}
\label{eq:ratio_R}
R \equiv \frac{\textrm{area of 1$\sigma$ allowed region obtained from the Fisher matrix analysis}}{\textrm{area of 1$\sigma$ allowed region obtained from the $\chi^2$ analysis}}.
\end{equation}
We expect $R$ to become unity when the prediction from the Fisher
matrix has a good agreement with that from the $\chi^2$ analysis.  In
Fig.~\ref{fig:area_single_power}, we show the ratio $R$ as a function
of SNR for different fiducial values of the spectral index, $n_{\rm
  GW}= -3, 0$ and $3$. Here raising the SNR is equivalent to
increase the fiducial amplitude $\Omega_{\rm GW\ast}$. Although the
tendency changes depending on the fiducial values of $n_{\rm GW}$, we
find that $R$ is close to unity when SNR is larger than $\sim 5$,
where we expect that we can safely adopt the Fisher matrix analysis.

\begin{figure}
  \begin{center}
    \includegraphics[width=6.4in]{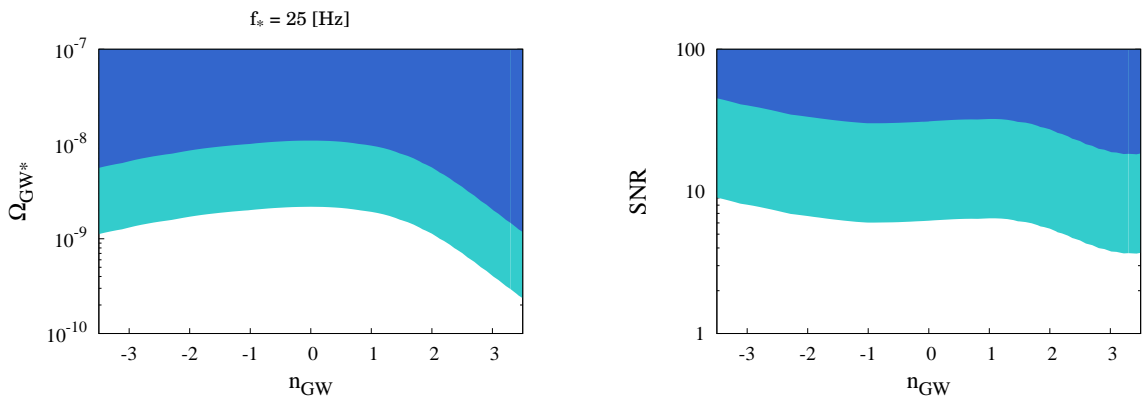}
  \end{center}
  \vspace{-5mm}
  \caption{Parameter space where $n_{\rm GW}$ can be determined with
    $\sigma_{n_{\rm GW}} < 0.1$ (blue) and $< 0.5$ (light blue) in the
    $n_{\rm GW}$--$\Omega_{\rm GW\ast}$ (left panel) and the
    $n_{\rm GW}$--SNR (right panel) planes.  Here, $\Omega_{\rm
      GW\ast}$ is marginalized over.  In the left panel, the reference
    frequency is taken at $f_\ast=25$~Hz.}
  \label{fig:error_single}
\end{figure}

Now we discuss to what extent we can determine the spectral index
$n_{\rm GW}$ in future GW experiments.  In general, the amplitude of
GW strongly depends on the values of model parameters, especially for
the cases of cosmological sources, while the spectral index $n_{\rm
  GW}$ does not, so that $n_{\rm GW}$ can be used to discriminate
sources of the GW background. Therefore, the accuracy of the
measurement of $n_{\rm GW}$ is of great interest.  In
Fig.~\ref{fig:error_single}, we show the parameter space where we can
determine the value of $n_{\rm GW}$ with an accuracy of
$\sigma_{n_{\rm GW}}<0.1$ (blue) and $<0.5$ (light blue) using
aLIGO-aVirgo-KAGRA.   It should be noted that when we show the result in the
$n_{\rm GW}$--$\Omega_{\rm GW\ast}$ plane, the shape of the contours
depends on $f_\ast$.  For example, if one takes the reference
frequency of $f_\ast \gg 25$ Hz ($f_\ast \ll 25$ Hz), the blue- (red-)
tilted spectrum cannot be probed.  On the other hand, when we take SNR
as the vertical axis instead of $\Omega_{\rm GW\ast}$, it does not
depend on $f_\ast$.  This is because $f_\ast$ is a redundant
parameter: Changing $f_\ast$ does not affect $n_{\rm GW}$ and can be
compensated by changing $\Omega_{\rm GW\ast}$ and hence gives the same
SNR (see Eqs.~\eqref{rhoIJ} and \eqref{rho}).  However, since the
$n_{\rm GW}$--$\Omega_{\rm GW\ast}$ plane is easier to understand
intuitively, we also show the results in the $n_{\rm
  GW}$--$\Omega_{\rm GW\ast}$ plane as well as in the $n_{\rm
  GW}$--SNR plane.

From the left panel, we find that, for $\Omega_{\rm GW\ast} > 10^{-8}$
and $2 \times 10^{-9}$, the spectral index $n_{\rm GW}$ can be
respectively determined with the accuracy of $\sigma_{n_{\rm GW}}<0.1$
and $<0.5$ for all range of $n_{\rm GW}$ shown in the figure.  We can
also notice that, when the spectrum is more tilted, particularly when
it is blue-tilted (i.e., $n_{\rm GW} >0$), $n_{\rm GW}$ can be better
probed compared to the scale-invariant spectrum (i.e., $n_{\rm GW} =
0$) for a fixed $\Omega_{\rm GW\ast}$.  This is because the
blue-tilted case is detectable with higher SNR for fixed $\Omega_{\rm
  GW\ast}$ and $f_\ast$, as seen in Fig.~\ref{fig:SN_Omega_n_single}

We would like to note that the value of SNR changes proportional to
$\Omega_{\rm GW\ast}$ as one can find by substituting
Eq.~\eqref{eq:broken_power} to Eq.~\eqref{rhoIJ}, so the result here
just scales as SNR~$\propto\Omega_{\rm GW\ast}$ when one takes a
different fiducial value of $\Omega_{\rm GW\ast}$ as long as the
weak-signal approximation is valid.  The same holds for the Fisher
matrix prediction \cite{Seto:2005qy}. Thus, in the following results,
the expected errors on the parameters can be scaled as $\sigma_{n_{\rm
    GW1,2}}\propto\Omega_{\rm GW\ast}^{-1}\propto {\rm SNR}^{-1}$.

\subsection{Broken power-law case}
\label{subsec:broken}

\begin{figure}
  \begin{center}
    \includegraphics[width=6.4in]{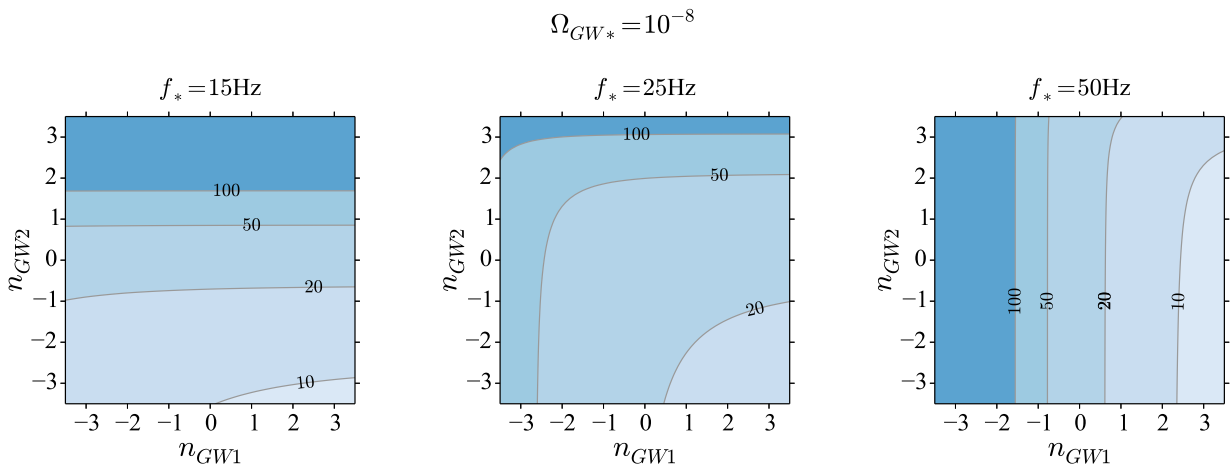}
  \end{center} 
  \vspace{-5mm}
  \caption{Detectability of the GWs for the broken power-law
    case. The contour plot of SNR in the $n_{\rm GW1}$--$n_{\rm GW2}$
    plane for $\Omega_{\rm GW\ast} =10^{-8}$ is shown for different
    reference frequency $f_\ast = 15~{\rm Hz}$ (left), $25~{\rm Hz}$
    (middle) and $50~{\rm Hz}$ (right).}
  \label{fig:detectability_broken}
\end{figure}

Next, we discuss the cases where we adopt the broken power-law form
Eq.~\eqref{eq:broken_power} as templates.  First, in
Fig.~\ref{fig:detectability_broken}, we show contours of SNR for
several values of $f_\ast$ by fixing $\Omega_{\rm GW\ast} =10^{-8}$ in
the $n_{\rm GW1}$--$n_{\rm GW2}$ plane.  One can notice that, the
dependence on $n_{\rm GW1}$ and $n_{\rm GW2}$ change depending on the
reference frequency. For example, the contours for the case of $f_\ast
= 50$~Hz are nearly vertical and $n_{\rm GW2}$ is irrelevant.  This is
because the number of frequency bands which is sensitive to $n_{\rm
  GW2}$ becomes smaller when $f_\ast$ is taken at higher frequency.  A
similar argument holds for the case of $f_\ast = 15$~Hz in which
$n_{\rm GW1}$ does not affect much the value of SNR and hence the
contours are almost horizontal along the axis of $n_{\rm GW1}$.

\begin{figure}
  \begin{center}
    \includegraphics[width=6.0in]{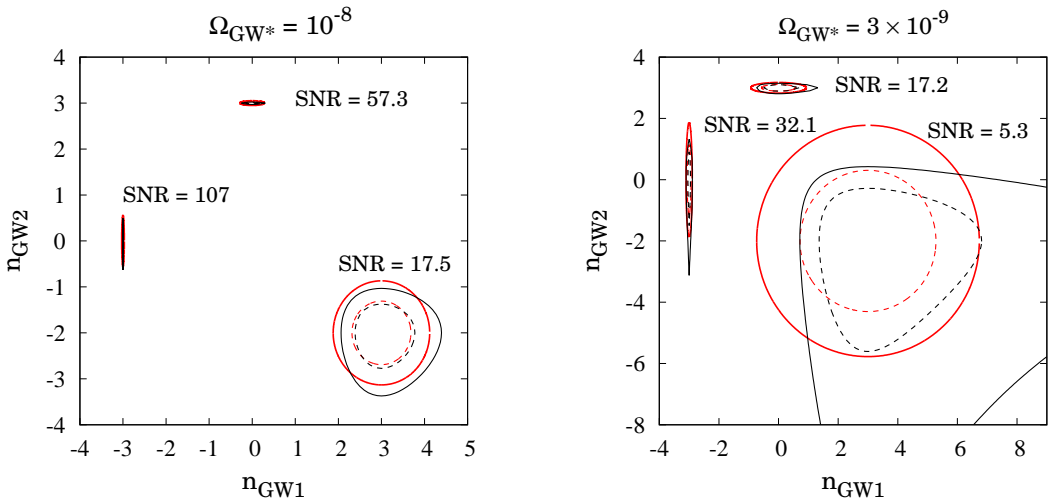}
  \end{center}
  \vspace{-10mm}
  \caption{1$\sigma$ (dashed line) and 2$\sigma$ (solid line) expected
    allowed region from the Fisher matrix (red) and the $\chi^2$
    (black) analyses in the $n_{\rm GW1}$ -- $n_{\rm GW2}$ plane for a
    broken power-law case.  Different fiducial parameter cases
    $(n_{\rm GW1}, n_{\rm GW2}) = (3,-2), (0,3)$ and $(-3,0)$ are
    shown with the values of SNR for each case.  Two panels show
    different fiducial value of $\Omega_{\rm GW \ast}=10^{-8}$ (left)
    and $3\times 10^{-9}$ (right).  The fiducial reference frequency
    is taken to be $f_\ast=25$~Hz.  Note that $\Omega_{\rm GW\ast}$
    and $f_\ast$ are not marginalized in this figure.}
  \label{fig:const_nGW12}
\end{figure}

Next, in Fig.~\ref{fig:const_nGW12}, we show examples of parameter
constraints in the $n_{\rm GW1}$--$n_{\rm GW2}$ plane assuming the
aLIGO-aVirgo-KAGRA observation.  Here, $\Omega_{\rm GW\ast}$ and
$f_\ast$ are not marginalized.  In the same way as
Fig.~\ref{fig:const_Omega_nGW_single}, we compare predictions made by
the $\chi^2$ and Fisher analyses. We show different sets of fiducial
parameters $(n_{\rm GW1}, n_{\rm GW2})=(3,-2)$, $(0,3)$ and $(-3,0)$,
with different fiducial amplitude $\Omega_{\rm GW\ast} = 10^{-8}$
(left) and $\Omega_{\rm GW\ast} = 3\times 10^{-9}$ (right).  The value
of SNR for each fiducial parameters is also shown in the figure.  As
seen in Fig.~\ref{fig:const_Omega_nGW_single}, the smaller SNR, the
more significant the deviation of the shape of the allowed region
between the $\chi^2$ and Fisher analyses, which again indicates that
the Fisher matrix analysis does not well describe expected constraints
when SNR is small.

\begin{figure}
  \begin{center}
    \includegraphics[width=3.0in]{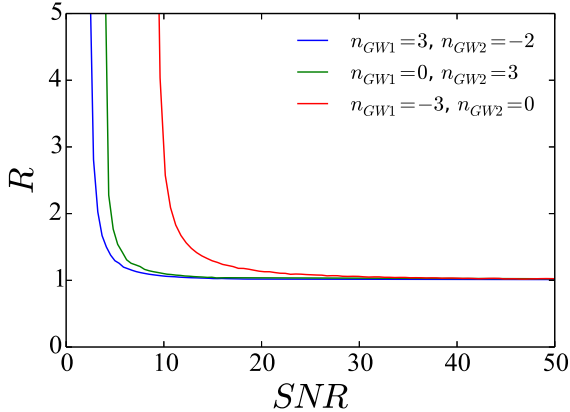}
  \end{center}
  \vspace{-5mm}
  \caption{Ratio $R$ (defined in Eq.~\eqref{eq:ratio_R}) as a function
    of SNR.  Cases with $(n_{\rm GW1}, n_{\rm GW2}) = (3,-2), (0,3)$
    and $(-3,0)$ are shown.  }
  \label{fig:ratio_R_broken_power}
\end{figure}

\begin{figure}
  \begin{center}
    \includegraphics[width=4.5in]{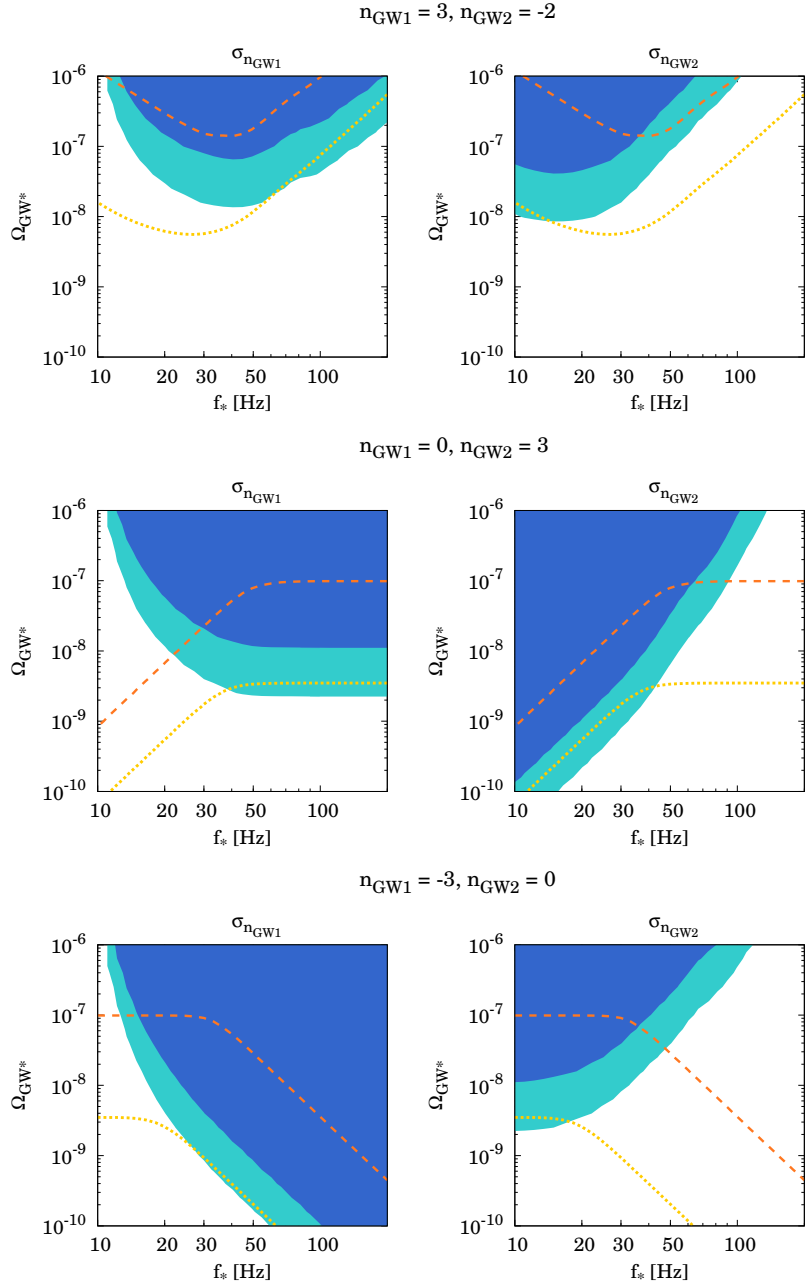}
  \end{center}
  \vspace{-10mm}
  \caption{Regions with $\sigma_{n_{\rm GW1}}< 0.1$ (blue) and $ <0.5$
    (light blue) are shown in the left panels, and those for
    $\sigma_{n_{\rm GW2}}<0.1$ and $<0.5$ are shown in the right
    panels in the $f_\ast$--$\Omega_{\rm GW\ast}$ parameter plane.
    The fiducial values of $(n_{\rm GW1}, n_{\rm GW2}) = (3,-2),
    (0,3)$ and $(-3,0)$ (from top to bottom) are shown. The region
    above the orange dashed line should be accessible with SNR $> 2$
    with the aLIGO O1 sensitivity, and the yellow dotted line
    indicates the accessible region by aLIGO-aVirgo-KAGRA with SNR
    $>10$.}
  \label{fig:error_broken_power}
\end{figure}

\begin{figure}
  \begin{center}
    \includegraphics[width=3.2in]{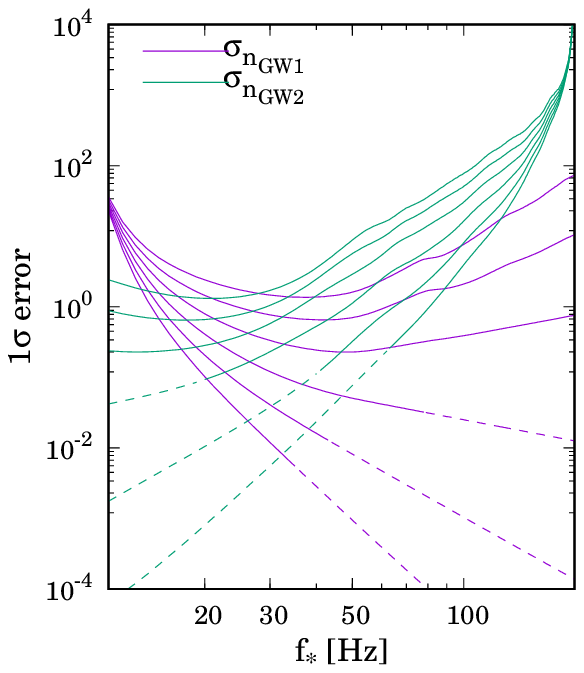}
  \end{center}
  \vspace{-5mm}
  \caption{Plots of $1\sigma$ errors of the spectral indices,
    $\sigma_{n_{\rm GW1}}$ and $\sigma_{n_{\rm GW2}}$, as a function
    of $f_\ast$.  Purple lines depict $\sigma_{n_{\rm GW1}}$ for the
    cases with $n_{\rm GW1} = 5, 3, 1, -1, -3, -5$ from top to bottom
    and $n_{\rm GW2} = -4$ fixed.  Green lines depict $\sigma_{n_{\rm
        GW2}}$ for the cases with $n_{\rm GW2} = 5, 3, 1, -1, -3, -5$
    from bottom to top and $n_{\rm GW1} = 4$ fixed.  Dashed lines
    correspond to the parameter regions which should be detectable
    with SNR $>2$ with the aLIGO O1 sensitivity.  The fiducial value
    of $\Omega_{\rm GW\ast}$ is fixed to be $10^{-8}$.  }
  \label{fig:error_broken_power_ref_scale_dep}
\end{figure}

\begin{figure}
  \begin{center}
    \includegraphics[width=6.4in]{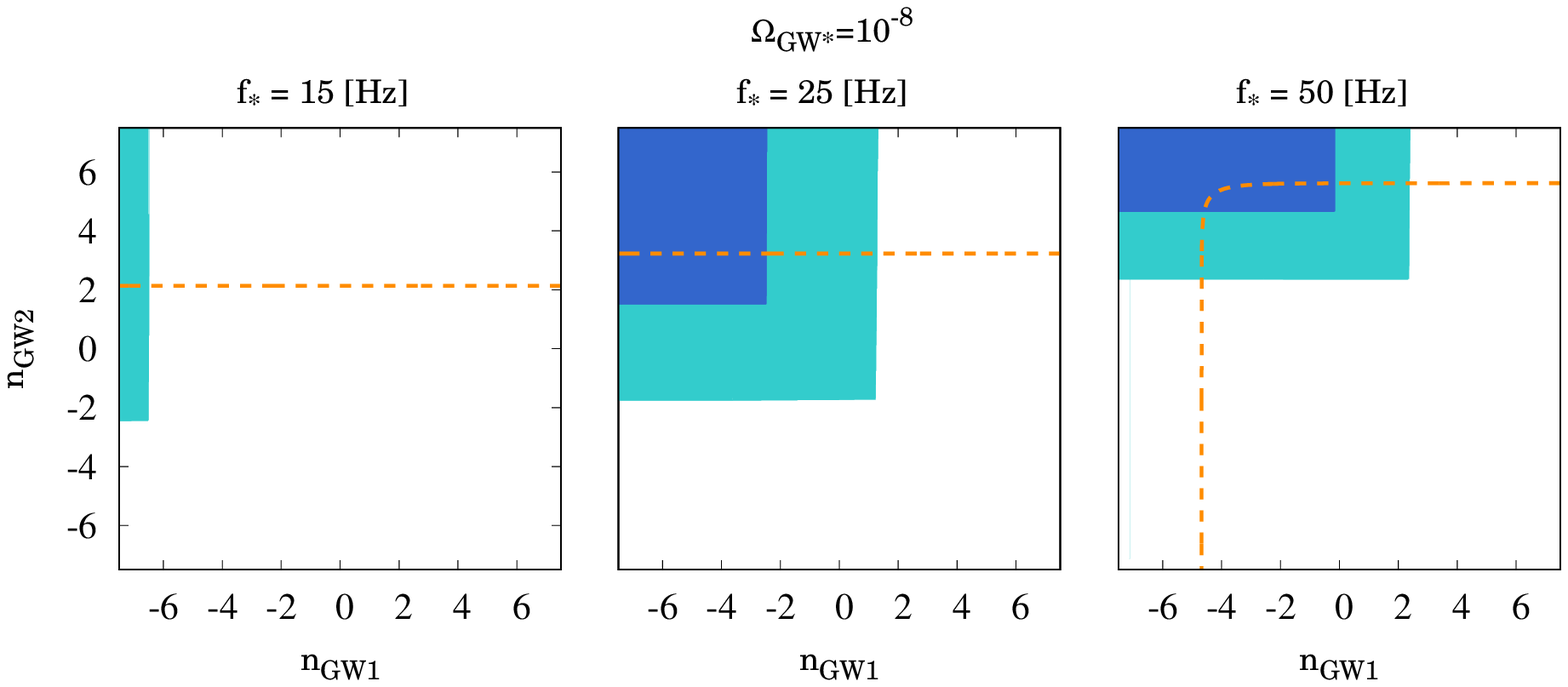}
  \end{center}
  \vspace{-10mm}
  \caption{Regions satisfying both of $\sigma_{n_{\rm GW1}} <0.1~(0.5)
    $ and $\sigma_{n_{\rm GW2}} <0.1~(0.5) $ are shown with blue
    (light blue) for the cases with $f_\ast = 15~{\rm Hz}$ (left),
    $25~{\rm Hz}$ (middle) and $50~{\rm Hz}$ (right).  The amplitude
    is fixed as $\Omega_{\rm GW\ast}=10^{-8}$.  Region above the
    orange dashed line should be accessible with SNR $>2$ with the
    aLIGO O1 sensitivity.  }
     \label{fig:error_broken_power_2}
\end{figure}

To quantify the validity of the Fisher matrix analysis, we again plot
the ratio $R$ defined in Eq.~\eqref{eq:ratio_R} for some sets of
$(n_{\rm GW1}, n_{\rm GW2})$ in Fig.~\ref{fig:ratio_R_broken_power}.
We fix the reference frequency at $f_\ast =25$ Hz.  In the same way as
single power-law case, when the value of SNR is larger (say SNR $\simg
10$), the ratio $R$ approaches unity, which means that the Fisher
matrix analysis gives a good estimate.  Note that the case of $(n_{\rm
  GW1}, n_{\rm GW2})=(-3,0)$ is exceptional because in this case, the
error contour is so elongated that the shape is almost one-dimensional
as seen from Fig.~\ref{fig:const_nGW12} and the area ratio $R$ may not
be a good indicator to check the validity of the Fisher matrix in this
case.  We also note here that the value of $R$ also depends on
$f_\ast$.  When we take the reference frequency away from $f_\ast
=25$Hz at which aLIGO O5 is most sensitive, the uncertainty of $n_{\rm
  GW1}$ or $n_{\rm GW2}$ gets larger and the line would shift to the
right.

In Fig.~\ref{fig:error_broken_power}, the parameter space where we can
determine $n_{\rm GW1}$ and $n_{\rm GW2}$ with $\sigma_{n_{\rm GW1}},
\sigma_{n_{\rm GW2}}<0.1$ and $0.5$ for different fiducial values of
$\Omega_{\rm GW\ast}$ and $f_\ast$ are shown in the
$f_\ast$--$\Omega_{\rm GW\ast}$ plane, which represents how precisely
we can determine the spectral indices with aLIGO-aVirgo-KAGRA.  Here
and in the following figures, Figs.~\ref{fig:error_broken_power},
\ref{fig:error_broken_power_ref_scale_dep} and
\ref{fig:error_broken_power_2}, other parameters are marginalized over
in the Fisher analysis. The orange dashed line in the figure describes
SNR $=2$ with the sensitivity of aLIGO O1 run.  Thus, in the region
above the orange dashed line, the GW background should be detectable
with the O1 sensitivity if we perform a stochastic GW background
search using broken power-law templates.  We also show the region
which can be accessible by aLIGO-aVirgo-KAGRA with SNR $>10$, whose
lower bound is indicated by the yellow dotted line.  From the left
bottom panel of Fig.~\ref{fig:error_broken_power}, one can easily
notice that for larger values of $f_\ast$, negative $n_{\rm GW1}$ can
be determined with high accuracy.  This is because a broad range of
the spectrum with $n_{\rm GW1}$ dependence is well inside the
sensitivity curve when the reference frequency is high.  Roughly
speaking, errors of $n_{\rm GW1}$ get smaller when the experiment can
measure the spectrum with $n_{\rm GW1}$ dependence in broad range of
frequencies as large SNR is obtained by summing up contributions from
each frequency bin.  The same argument holds for $n_{\rm GW2}$.

The $f_\ast$ dependence can be clearly seen in
Fig.~\ref{fig:error_broken_power_ref_scale_dep} where 1$\sigma$ error
is plotted as a function of $f_\ast$ for $n_{\rm GW1} =5, 3, 1, -1,
-3, -5$ with $n_{\rm GW2} = -4$ being fixed (and $n_{\rm GW2} = 5, 3,
1, -1, -3, -5$ with $n_{\rm GW1} = 4$.).  We can see the tendency that
the error of $n_{\rm GW1}$ improves when $n_{\rm GW1}$ is more
negative and $f_\ast$ is larger, while the error of $n_{\rm GW2}$
improves when $n_{\rm GW2}$ is more positive and $f_\ast$ is smaller,
because they give larger SNR for fixed $\Omega_{\rm GW\ast}$.  When
$n_{\rm GW1}=5$ ($n_{\rm GW2}=-5$), we see that the error does not
improve even when $f_\ast$ is high (low). This is because the spectral
slope is so steep that the spectrum goes below the sensitivity curve
quickly at low (high) frequencies and cannot increase the SNR.

In Fig.~\ref{fig:error_broken_power_2}, the regions satisfying both
$\sigma_{n_{\rm GW1}} <0.1~(0.5)$ and $\sigma_{n_{\rm GW2}}
<0.1~(0.5)$ are depicted with dark blue (light blue) in the $n_{\rm
  GW1}$--$n_{\rm GW2}$ plane for $f_\ast = 15$~Hz, $25$~Hz and $50$~Hz
with $\Omega_{\rm GW\ast} = 10^{-8}$ being fixed.  The right edge of
the rectangle is determined by $\sigma_{n_{\rm GW1}} <0.1~(0.5)$ and
the lower edge of the rectangle is determined by $\sigma_{n_{\rm GW2}}
<0.1~(0.5)$.  The GW background should be detectable with aLIGO O1 run
by SNR $>2$ in the region above the orange dashed line.  As we would
naively expect, we see that the GW spectrum can be well probed when
the GW spectrum is convex downward (i.e., a negative $n_{\rm GW1}$ and
a positive $n_{\rm GW2}$) since they give a large SNR for a fixed
$\Omega_{\rm GW\ast}$.  We can also see the tendency that, when we
take a smaller $f_\ast$ (such as the case with $f_\ast = 15~{\rm
  Hz}$), $n_{\rm GW2}$ can be easily probed.  For larger $f_\ast$, we
can better determine $n_{\rm GW1}$.  Note that the case with $f_\ast =
15~{\rm Hz}$ has a small parameter space where we can determine
$n_{\rm GW1}$, because we have the low frequency cut off at $10$~Hz
and the frequency range where $n_{\rm GW1}$ can be well probed is very
narrow.

\subsection{Expected constraints from DECIGO}
\label{sec:DECIGO}

\begin{figure}
  \begin{center}
    \includegraphics[width=6.4in]{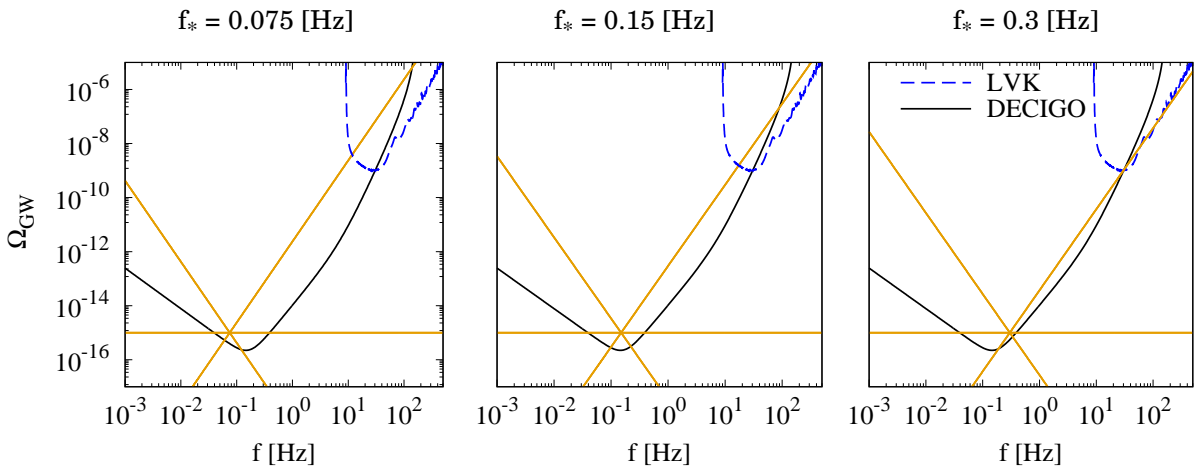}
  \end{center}
  \vspace{-10mm}
  \caption{Sensitivity curve for DECIGO (black solid). For comparison, we
    also plot the sensitivity curve of aLIGO-aVirgo-KAGRA (LVK) (blue dashed) as
    well as the spectra with $n_{\rm GW1}$ and $n_{\rm GW2}$ being
    assumed to be $-3, 0$ and $3$ (yellow). We take $f_\ast =
    0.075~{\rm Hz}$ (left), $0.15~{\rm Hz}$ (middle) and $0.3~{\rm
      Hz}$ (right). The amplitude at $f_\ast$ is fixed to be
    $\Omega_{\rm GW\ast} = 10^{-15}$. }
     \label{fig:spectrum_decigo}
\end{figure}

\begin{figure}
  \begin{center}
    \includegraphics[width=3.2in]{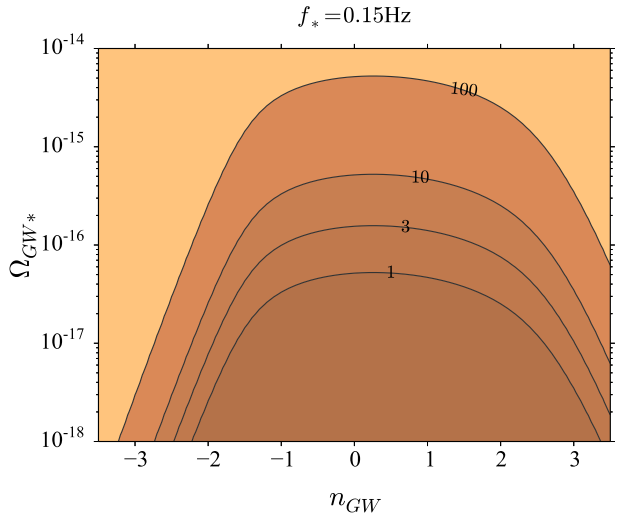}
  \end{center}
  \vspace{-5mm}
  \caption{Contours of SNR in the $n_{\rm GW}$--$\Omega_{\rm GW\ast}$
    plane for a single power-law case expected from DECIGO.  We take
    the reference frequency as $f_\ast = 0.15~{\rm Hz}$, which is
    the most sensitive frequency for DECIGO.}
  \label{fig:SNR_single_power_DECIGO}
\end{figure}

\begin{figure}
  \begin{center}
    \includegraphics[width=6.4in]{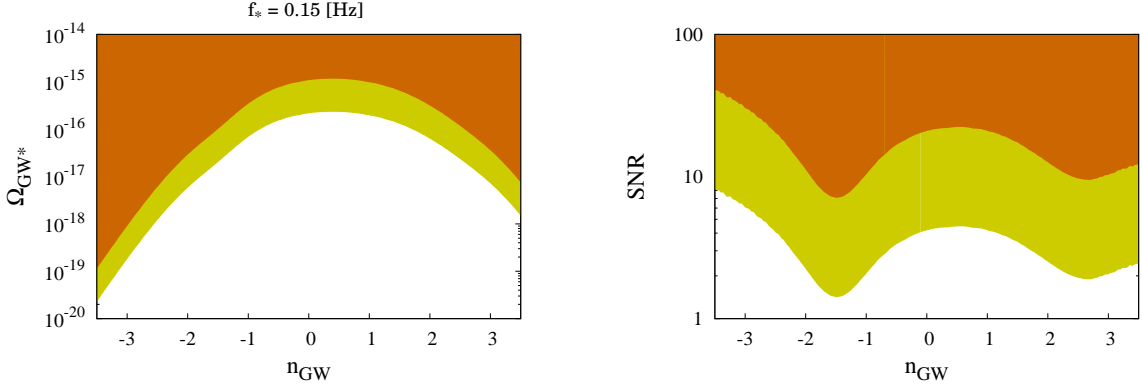}
  \end{center}
  \vspace{-5mm}
  \caption{Parameter space where $n_{\rm GW}$ can be determined in
    DECIGO with $\sigma_{n_{\rm GW}} < 0.1$ (orange) and $< 0.5$
    (yellow) in the $n_{\rm GW}$--$\Omega_{\rm GW\ast}$ (left) and
    the $n_{\rm GW}$--SNR (right) planes.  Here, $\Omega_{\rm GW\ast}$
    is marginalized over.  In the left panel, the reference frequency
    is taken at $f_\ast=0.15$~Hz.}
  \label{fig:error_DECIGO}
\end{figure}

\begin{figure}
  \begin{center}
    \includegraphics[width=6.4in]{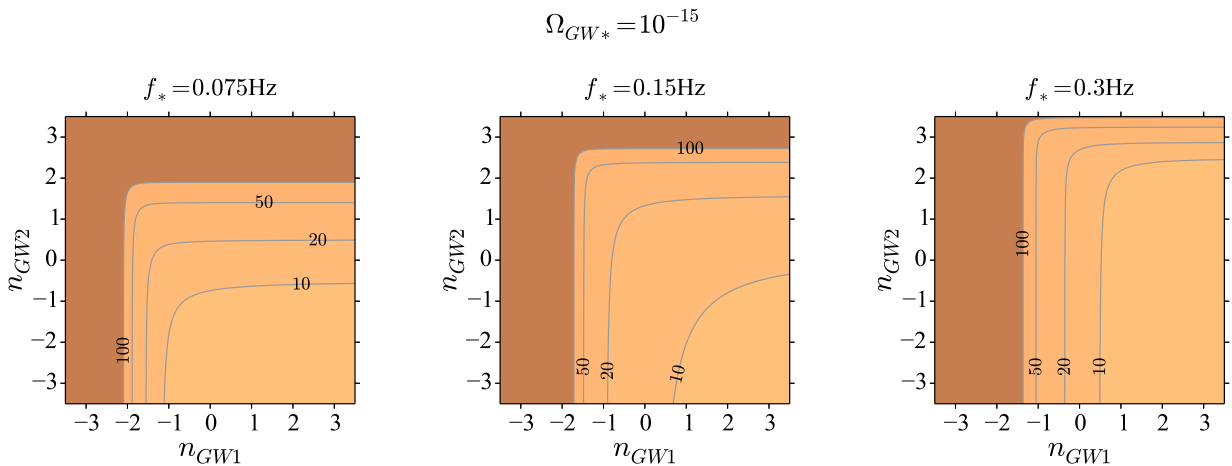}
  \end{center}
  \vspace{-5mm}
  \caption{
    Detectability of DECIGO for the broken power-law case. The contour
    plot of SNR in the $n_{\rm GW1}$--$n_{\rm GW2}$ plane for
    $\Omega_{\rm GW\ast} =10^{-15}$ is shown for different reference
    frequency $f_\ast = 0.075~{\rm Hz}$ (left), $0.15~{\rm Hz}$ (middle)
    and $0.3~{\rm Hz}$ (right).}  
  \label{fig:detectability_broken_DECIGO}
\end{figure}

\begin{figure}
  \begin{center}
    \includegraphics[width=4.3in]{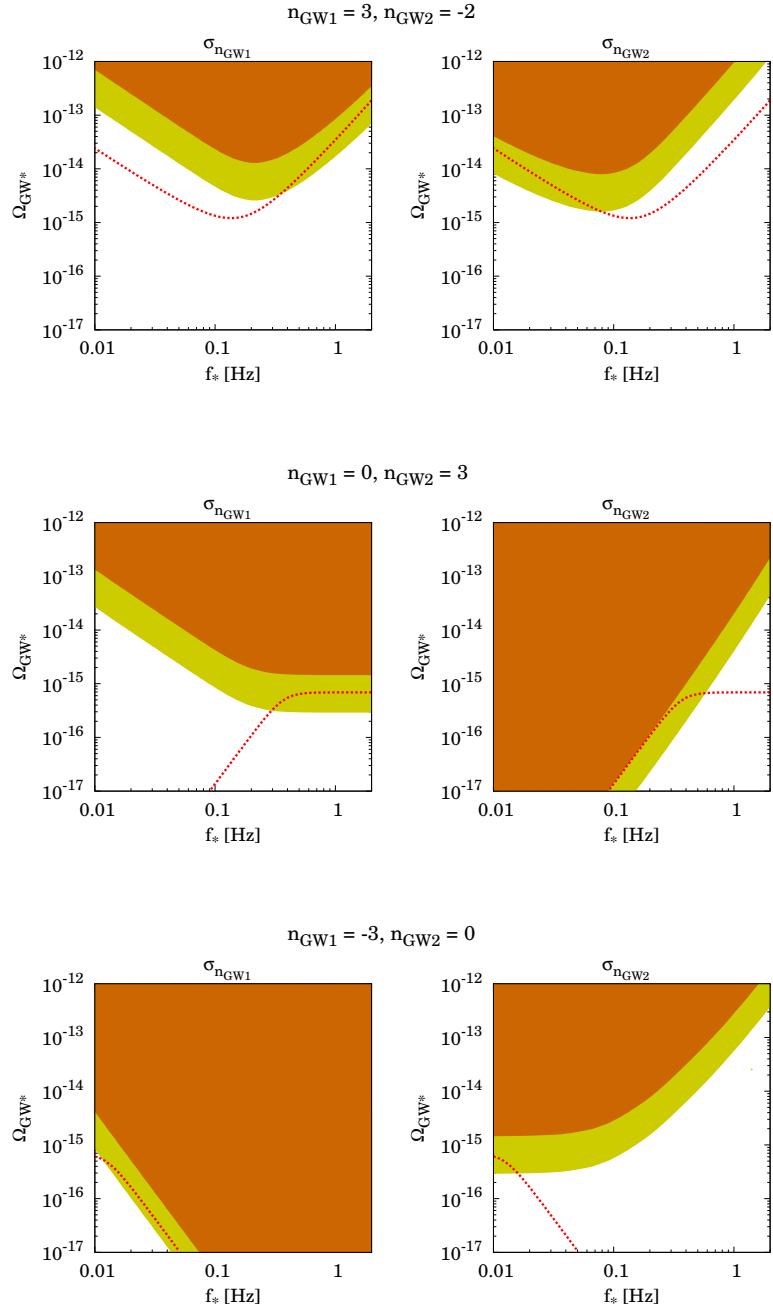}
  \end{center}
  \vspace{-10mm}
  \caption{ Errors on spectral indices expected in DECIGO are shown in
    the $\Omega_{\rm GW\ast}$--$f_\ast$ plane. The parameter space
    satisfying $\sigma_{n_{\rm GW1}} < 0.1$ (orange) and $ <0.5$
    (yellow) are presented in the left panels, and those for
    $\sigma_{n_{\rm GW2}}$ are in the right panels.  Cases with
    $(n_{\rm GW1}, n_{\rm GW2}) = (3,-2), (0,3)$ and $(-3,0)$ (from
    top to bottom) are shown.  The red dotted line represents the accessible
    region by DECIGO with SNR $>10$. }
  \label{fig:n_GW1_n_GW2_error}
\end{figure}

\begin{figure}
  \begin{center}
    \includegraphics[width=3.2in]{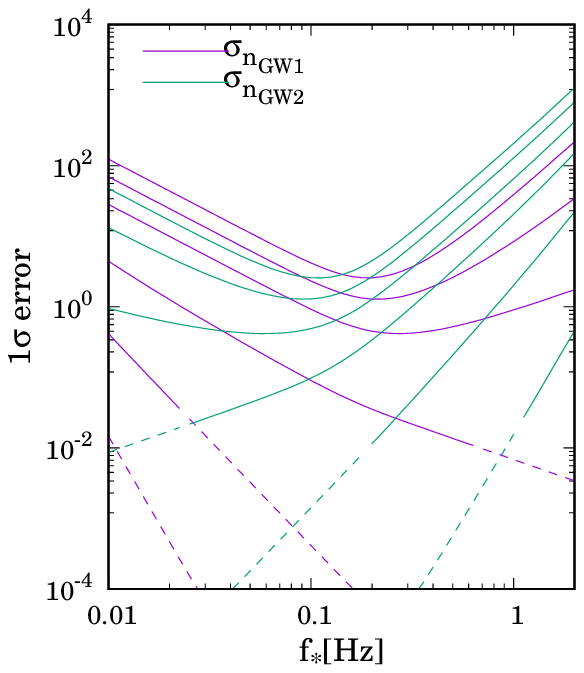}
  \end{center}
  \vspace{-5mm}
  \caption{Plots of $1\sigma$ errors of the spectral indices,
    $\sigma_{n_{\rm GW1}}$ and $\sigma_{n_{\rm GW2}}$, as a function
    of $f_\ast$.  Purple lines depict $\sigma_{n_{\rm GW1}}$ for the
    cases with $n_{\rm GW1} = 5, 3, 1, -1, -3, -5$ from top to bottom
    and $n_{\rm GW2} = -4$ fixed.  Green lines depict $\sigma_{n_{\rm
        GW2}}$ for the cases with $n_{\rm GW2} = 5, 3, 1, -1, -3, -5$
    from bottom to top and $n_{\rm GW1} = 4$ fixed.  Dashed lines
    correspond to the parameter range where the weak-signal
    approximation breaks down with ${\rm SNR} > 100$. The fiducial value of
    $\Omega_{\rm GW\ast}$ is fixed to be $\Omega_{\rm GW\ast}=10^{-15}$.  }
  \label{fig:sigma_nGW1_nGW2_f_ast_DECIGO}
\end{figure}

\begin{figure}
  \begin{center}
    \includegraphics[width=6.4in]{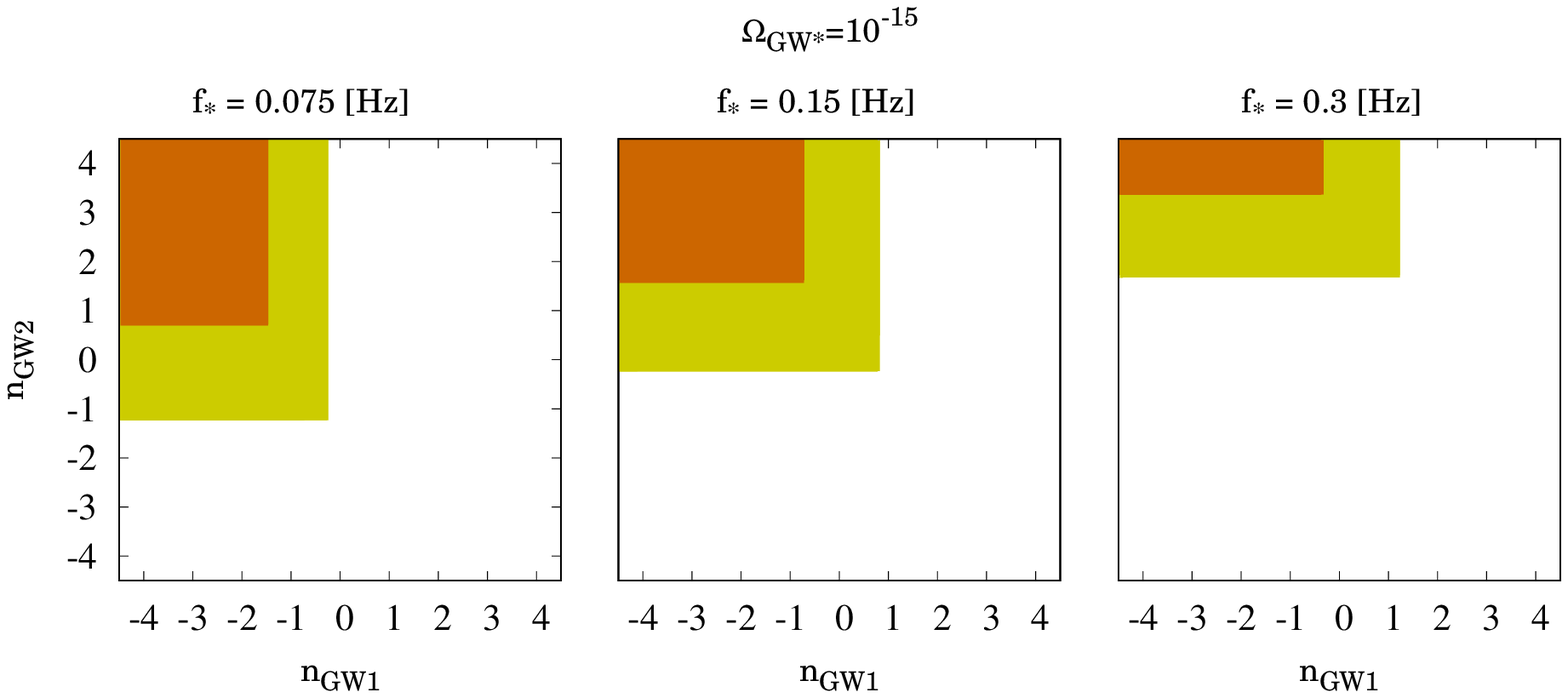}
    \includegraphics[width=6.4in]{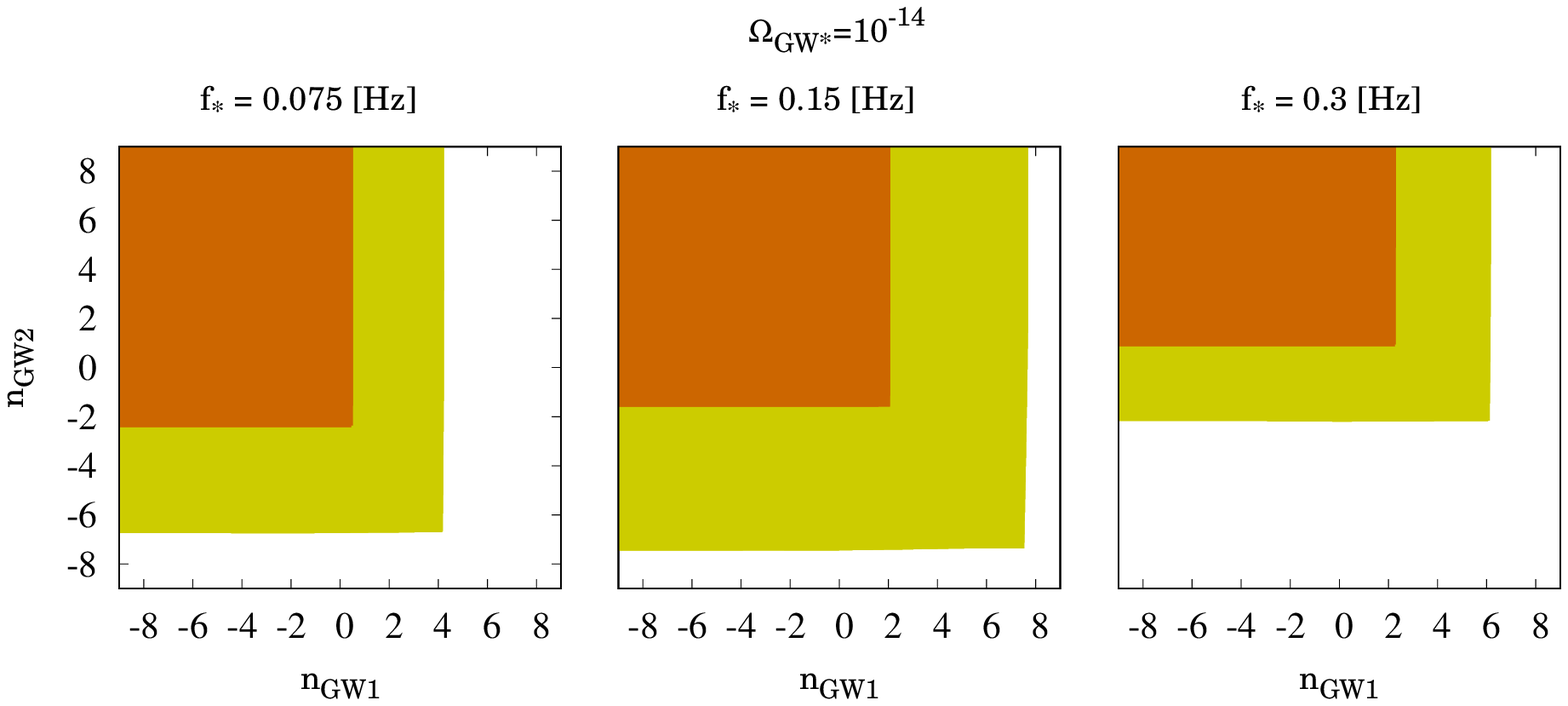}
  \end{center}
  \vspace{-5mm}
  \caption{Regions satisfying both of $\sigma_{n_{\rm GW1}}
    <0.1~(0.5)$ and $\sigma_{n_{\rm GW2}} <0.1~(0.5)$ expected from
    DECIGO are shown with orange (yellow) for the cases with $f_\ast =
    0.075~{\rm Hz}$ (left), $0.15~{\rm Hz}$ (middle) and $0.3~{\rm
      Hz}$ (right).  The fiducial amplitude is fixed to be $\Omega_{\rm
      GW\ast}=10^{-15}$ (upper panels) and $\Omega_{\rm GW\ast}=10^{-14}$
    (lower panels). }
  \label{fig:error_broken_power_DECIGO}
\end{figure}

We repeat the same analysis by assuming the specification of DECIGO in
this section.  The results are shown in
Figs.~\ref{fig:spectrum_decigo}--\ref{fig:error_broken_power_DECIGO}.
The tendencies are almost the same as those obtained for
aLIGO-aVirgo-KAGRA, however, as seen from
Fig.~\ref{fig:spectrum_decigo} where the sensitivity curve for DECIGO
is shown, there are two important differences: (i)~the frequency range
sensitive to the signal, (ii)~the sensitivity to $\Omega_{\rm
  GW\ast}$.  DECIGO is sensitive to the frequency of $f \sim 0.1~{\rm
  Hz}$ and the sensitivity curve reaches  $\Omega_{\rm GW\ast} \sim
2 \times 10^{-16}$, which is about 7 orders of magnitude better than
aLIGO-aVirgo-KAGRA.  

Due to these differences, we take the reference frequency as (or close
to) $f_\ast = 0.15~{\rm Hz}$ in
Figs.~\ref{fig:SNR_single_power_DECIGO},
\ref{fig:error_DECIGO} and
\ref{fig:detectability_broken_DECIGO}, while $f_\ast$ taken to be
${\cal O}(10)$~Hz for aLIGO-aVirgo-KAGRA.  Regarding the amplitude at
the reference frequency $\Omega_{\rm GW\ast}$, we take it to be
$\Omega_{\rm GW\ast} = 10^{-15}$ in
Fig.~\ref{fig:sigma_nGW1_nGW2_f_ast_DECIGO} and 
\ref{fig:error_broken_power_DECIGO}, while we mainly used $\Omega_{\rm
  GW\ast} = 10^{-8}$ for aLIGO-aVirgo-KAGRA.  Also since we assume the
wide frequency range $10^{-3}$ -- $10^2$~Hz for DECIGO, while $10$ --
$200$~Hz for aLIGO-aVirgo-KAGRA, the result tends to be more sensitive
to the change in the values of the spectral indices $n_{\rm GW1}$ and
$n_{\rm GW2}$, since the amplitude of $\Omega_{\rm GW}(f)$ with tilt
changes a lot when the frequency range is wide.

First, we consider the single power-law case in the same way as in
Sec.  \ref{subsec:single}.  In Fig.\ref{fig:SNR_single_power_DECIGO},
we estimate SNR for a single power-law case with $f_\ast=0.15$~Hz in
the $n_{\rm GW}$--$\Omega_{\rm GW\ast}$ plane.  The figure clearly
shows that DECIGO can probe much larger parameter space as it can
detect GWs of $\Omega_{\rm GW\ast}\sim 10^{-15}$ with SNR $>10$ for any value of
$n_{\rm GW}$.  In Fig.~\ref{fig:error_DECIGO}, we show the parameter
space where we can determine $n_{\rm GW}$ with $\sigma_{n_{\rm
    GW}}<0.1$ (orange) and $<0.5$ (yellow) using DECIGO. We find that
SNR~$>10$ is required to determine $n_{\rm GW}$, which is similar to
the aLIGO-aVirgo-KAGRA case.  If the stochastic GW background is
detected by aLIGO-aVirgo-KAGRA, that would provide a prediction for
the frequency range of DECIGO.  For example, for $n_{\rm GW}=2/3$,
which is the power index of the background generated by compact binary
coalescence (see Sec. \ref{subsec:astrophysical}), the detection by
aLIGO-aVirgo-KAGRA (LVK) would imply the amplitude at DECIGO as
$\Omega_{\rm GW\ast}^{DECIGO}= \Omega_{\rm GW\ast}^{LVK}
\left(\frac{0.15}{25}\right)^{2/3} \simeq 5.9\times 10^{-11}
\left(\frac{\Omega_{\rm GW\ast}^{LVK}}{1.8\times 10^{-9}}\right) $,
which should be detected by DECIGO with high SNR\footnote{ The
  amplitude at the frequency of $10^{-2}$Hz would
  be $\Omega_{\rm GW\ast}^{LISA}=\Omega_{\rm
      GW\ast}^{LVK}\left(\frac{10^{-2}}{25}\right)^{2/3} \simeq
  9.8\times 10^{-12} \left(\frac{\Omega_{\rm GW\ast}^{LVK}}{1.8\times
    10^{-9}}\right)$, which is also expected to be detected by LISA \cite{Audley:2017drz}.}.
      Even for
$n_{\rm GW}=3$, the amplitude at DECIGO becomes
\begin{equation} 
\Omega_{\rm GW\ast}^{DECIGO}
= 3.9\times 10^{-16}
\left(\frac{\Omega_{\rm GW\ast}^{LVK}}{1.8\times 10^{-9}}\right)\left(\frac{0.15 {\rm Hz}}{25 {\rm Hz}}\right)^{n_{\rm GW}-3} ,
\end{equation} 
which would be detected by DECIGO as shown in Fig.~\ref{fig:error_DECIGO}. 
Therefore, the detection of the stochastic GW background by DECIGO would 
be a consistency check of the detection by aLIGO-aVirgo-KAGRA.

Next, we consider the broken power-law case in the same way as in
Sec. \ref{subsec:broken}.  In
Fig.~\ref{fig:detectability_broken_DECIGO}, we show the contours of
SNR for several values of $f_\ast$ fixing $\Omega_{\rm GW\ast}
=10^{-15}$ in the $n_{\rm GW1}$--$n_{\rm GW2}$ plane.  We can find the
same tendency with the case of aLIGO-aVirgo-KAGRA, but notice that the
assumed value of $\Omega_{\rm GW\ast}$ is much smaller here.  We can
also see that the SNR is more sensitive to the change of $n_{\rm GW1}$
and $n_{\rm GW2}$ because of the wide frequency range $10^{-3}$ --
$10^2$~Hz for DECIGO.  In Fig.~\ref{fig:n_GW1_n_GW2_error}, we show
the parameter space where the spectral indices can be determined with
$\sigma_{n_{\rm GW1}}<0.1$ and $<0.5$ (left) and $\sigma_{n_{\rm
    GW2}}<0.1$ and $<0.5$ (right) in the $\Omega_{\rm
  GW\ast}$--$f_\ast$ plane.  Here and in the following figures,
Figs.~\ref{fig:n_GW1_n_GW2_error},
\ref{fig:sigma_nGW1_nGW2_f_ast_DECIGO} and
\ref{fig:error_broken_power_DECIGO}, other parameters are marginalized
over in the analysis.  For $(n_{\rm GW1},n_{\rm GW2})=(3,-2)$, which
cannot be measured by aLIGO-aVirgo-KAGRA with $\Omega_{\rm
  GW\ast}=10^{-8}$, both indices can be measured with $\sigma_{n_{\rm
    GW1,2}}<0.1$ even with $\Omega_{\rm GW\ast}=10^{-13}$ for $0.03
{\rm Hz}\siml f_\ast\siml 0.4$ Hz.
  
Fig.~\ref{fig:sigma_nGW1_nGW2_f_ast_DECIGO} shows $f_\ast$ dependence
of 1$\sigma$ uncertainties of spectral indices, $\sigma_{n_{\rm GW1}}$
and $\sigma_{n_{\rm GW2}}$.  In both figures, we find the same
tendency with the case of aLIGO-aVirgo-KAGRA, but again the accessible
amplitude and the frequency are different.  Finally, in
Fig.~\ref{fig:error_broken_power_DECIGO}, we show parameter space
where both $\sigma_{n_{\rm GW1}}<0.1 (0.5)$ and $\sigma_{n_{\rm
    GW1}}<0.1 (0.5)$ are satisfied in the $n_{\rm GW1}$--$n_{\rm GW2}$
plane.  The upper panels may give impression that only small
parameter space can be probed, but this is just because the fiducial
amplitude assumed here, $\Omega_{\rm GW\ast}=10^{-15}$, is small. As
seen in the lower panels, for $\Omega_{\rm GW\ast}=10^{-14}$, almost
all the parameter space can be covered with DECIGO.

\section{Summary}
\label{sec:summary}

Since the first detection of the GWs from the merger of a black hole
binary \cite{Abbott:2016blz}, several GWs from the merger of binary
black holes and neutron stars have been detected.  We are now in the
stage of a possible detection of the stochastic GW background from
compact binary coalescence \cite{Abbott:2017xzg}.  As we have seen in
Sec.~\ref{sec:GW_source}, there are lots of sources of stochastic
GW background of cosmological origin as well as astrophysical ones.
Most of the spectra of stochastic GW background cannot be fitted by a
single power law, but rather they are better fitted by a broken
power law.

In this paper, we have demonstrated the use of the broken power-law
templates.  We have also calculated the expected constraints on the
parameters such as spectral indices by using the Fisher matrix
analysis for both single and broken power-law templates, assuming
the sensitivity of the future detector network of aLIGO-aVirgo-KAGRA and of
a future detector DECIGO.  For aLIGO-aVirgo-KAGRA, we have found
that the spectral index of a single power-law template can be measured
with $\sigma_{n_{\rm GW}}<0.1$ if $\Omega_{\rm GW\ast}>10^{-8}$ and
that two indices of a broken power-law template with $f_\ast=25$ Hz
(50 Hz) can be measured with an accuracy of $\sigma_{n_{\rm GW1,2}}<0.5$ for $n_{\rm
  GW1}\siml 1 (2)$ and $n_{\rm GW2}\simg -2 (2)$ for $\Omega_{\rm
  GW\ast}>10^{-8}$.  We have also estimated the required SNR in order
for the Fisher matrix analysis to provide an accurate estimate of the
parameters by comparing with the result from the $\chi^2$ analysis.

The accuracy would be improved significantly for DECIGO.  The spectral
index of a single power-law spectrum can be measured with
$\sigma_{n_{\rm GW}}<0.1$ even for $\Omega_{\rm GW\ast}>10^{-15}$.
For a broken power-law spectrum with $(n_{\rm GW1},n_{\rm
  GW2})=(3,-2)$, which cannot be measured by aLIGO-aVirgo-KAGRA with
$\Omega_{\rm GW\ast}=10^{-8}$, both indexes can be measured with
$\sigma_{n_{\rm GW1,2}}<0.1$ even with $\Omega_{\rm GW\ast}=10^{-13}$
for $0.03 {\rm Hz}\siml f_\ast\siml 0.4$ Hz.  With a possible
detection of the stochastic background by aLIGO-aVirgo-KAGRA, the
measurement by DECIGO could be used as a consistency check of the
spectrum of the background.

The spectral indices would be useful to narrow down the sources of the
background.  Furthermore, it may also be possible to discriminate
between a smooth background from cosmological sources and a discrete
``popcorn-type'' background such as the one from astrophysical sources
and the one from the smooth stochastic background from the early
universe sources (such as inflation) by measuring the non-Gaussianity
of the GW data streams \cite{Seto:2008xr} or by the anisotropies of
the spectrum \cite{Cusin:2018rsq}.  By combining this information with
the spectral indices studied in this paper, we can deepen our
understandings of the Universe through the stochastic GW background.

\acknowledgments TC would like to thank Takahiro Tanaka for useful
communications.  The authors are grateful to Ryusuke Jinno for useful
comments.  This work is partially supported by MEXT KAKENHI Grant
Number 15H05894 (TC), 15H05888 (TT), by JSPS KAKENHI Grant Number
17K14282 (SK), 15K05084 (TT), 17H01131 (TT), by the Career Development
Project for Researchers of Allied Universities (SK), and in part by
Nihon University (TC).

\end{document}